\documentclass[%
 aps,
 amsmath,amssymb,
 reprint,%
]{revtex4-1}

\draft 

\usepackage[latin1]{inputenc}
\usepackage{blindtext}
\usepackage{centernot}
\usepackage{graphicx}
\usepackage{amsmath,bbold}
\usepackage{times}
\usepackage{amssymb}
\usepackage{mathrsfs}
\usepackage{chemarr}
\usepackage{color}
\usepackage{url}
\usepackage{version}
\usepackage[hidelinks]{hyperref}
\usepackage{mwe,tikz}
\usepackage[percent]{overpic}
\usepackage{bm}
\usepackage[export]{adjustbox}
\definecolor{linkcolor}{rgb}{0,0,0.6} 
\usepackage{lipsum}
\usepackage[normalem]{ulem}

\usepackage{chngcntr}

\usepackage{mathtools}
\DeclarePairedDelimiter\abs{\lvert}{\rvert}

\newcommand{\niceVec}[2]{\vec{#1}^{\,#2}}
\newcommand{\nvecsub}[3]{\vec{#1}^{\,#2}_{\,#3}}
\newcommand{\mat}[1]{{\boldsymbol #1}}

\definecolor{safeGreen}{rgb}{0.0, 0.5, 0.0}              
\newcommand{\mycomment}[1]{[{\bf{\color{safeGreen}#1}]}} 

\newcommand{\myreply}[1]{{\bf[{\color{magenta}#1}]}}     

\newcommand{\TF}{T_\text{f}}

\newcommand{\nsub}{\mathcal{N}}
\newcommand{\nclus}{n}
\newcommand{\bonds}{\mathcal{B}}
\newcommand{\state}{\mathcal{S}}
\newcommand{\mfrac}{f_{\text{1}}}
\newcommand{\Ntot}{N_\text{tot}}

\newcommand{\ctot}{c_0}
\newcommand{\ctotMax}{c_0^\text{max}}
\newcommand{\css}{c_\text{ss}}

\newcommand{\discIndex}{N}

\newcommand{\kt}{k_\text{B}T}
\newcommand{\NA}{N_\text{A}}

\newcommand{\tzero}{t_0}
\newcommand{\lzero}{l_0}
\newcommand{\Lzero}{L_0}

\newcommand{\Ebind}{\epsilon_{11}}
\newcommand{\Ett}{\epsilon_\text{TT}}
\newcommand{\Etb}{\epsilon_\text{TB}}
\newcommand{\Stt}{\sigma_\text{TT}}
\newcommand{\Stb}{\sigma_\text{TB}}
\newcommand{\ep}{\epsilon_\text{p}}

\newcommand{\Eone}{\epsilon_{12}}
\newcommand{\Ethree}{\epsilon_{33}}

\begin{document}

\title{Markov State Model Approach to Simulate Self-Assembly}

\author{Anthony Trubiano}
\email[]{trubiano@brandeis.edu}
\author{Michael F. Hagan}
\email[]{hagan@brandeis.edu}
\affiliation{Martin Fisher School of Physics, Brandeis University, Waltham, Massachusetts 02454, USA}

\date{\today}

\begin{abstract}
Computational modeling of assembly is challenging for many systems because their timescales vastly exceed those accessible to simulations. This article describes the MultiMSM, which is a general framework that uses Markov state models (MSMs) to enable simulating self-assembly and self-organization on timescales that are orders of magnitude longer than those accessible to brute force dynamics simulations. In contrast to previous MSM approaches to simulating assembly, the framework describes simultaneous assembly of many clusters and the consequent depletion of free subunits or other small oligomers. The algorithm accounts for changes in transition rates as concentrations of monomers and intermediates evolve over the course of the reaction. Using two model systems, we show that the MultiMSM accurately predicts the concentrations of the full ensemble of intermediates on the long timescales required for reactions to reach equilibrium. Importantly, after constructing a MultiMSM for one system concentration, a wide range of other concentrations can be simulated without any further sampling. This capability allows for orders of magnitude additional speed up. In addition, the method enables highly efficient calculation of quantities such as free energy profiles, nucleation timescales, flux along the ensemble of assembly pathways, and entropy production rates. Identifying contributions of individual transitions to entropy production rates reveals sources of kinetic traps. The method is broadly applicable to systems with equilibrium or nonequilibrium dynamics, and is trivially parallelizable and thus highly scalable.
\end{abstract}

\maketitle 

\section{Introduction}
The self-assembly of basic subunits into larger, more complex structures is fundamental to life. Critical functions of cells and pathogens are performed by assembled structures such as the outer shells (capsids) of viruses \cite{Caspar1962, Mateu2013, Bruinsma2015, Perlmutter2015, Hagan2016, Twarock2018, Zandi2020, Hagan2021} or bacterial microcompartments \cite{Kerfeld2010,Tanaka2008,Rae2013,Bobik2015,Chowdhury2014,Kerfeld2016,Polka2016,Kerfeld2018,SliningerLee2017}, cytoskeletal filaments \cite{Banerjee2020, Cabeen2010, Pilhofer2013}, and ordered protein layers on bacteria exteriors \cite{Whitelam2010}. Self-assembly is also transforming nanotechnology, where designing synthetic building blocks that are preprogrammed to form particular structures is enabling scalable bottom-up synthesis of materials with desirable properties \cite{Garg2015, Beija2012, Ebbens2016, Mallory2018, Fan2011, Huh2020, Ke2012, Kraft2009, Sacanna2010, Sacanna2013, Yi2013, Wang2014, He2020, He2021, Chen2011a, Chen2011b, Zerrouki2008, Yan2013, Wolters2015,Tikhomirov2018, Oh2019, Ben2021, Kahn2022}. 

Since it is an inherently out-of-equilibrium process, understanding or designing  self-assembly processes requires detailed knowledge of assembly intermediates and the dynamical transitions among them. Computational modeling is an essential tool for revealing such assembly pathways, since most intermediates are too transient to characterize in experiments. However, simulating assembly at experimentally relevant conditions is intractable for many models, since assembled structures are much larger than their components and form on timescales that are orders of magnitude beyond computational limitations. This article describes a Markov state model (MSM) framework that overcomes this limitation for a broad array of self-assembly and self-organization systems. The algorithm reduces computational times by orders of magnitude while describing the time-dependent concentrations of subunits and the complete ensemble of assembly intermediates and products. This capability enables dynamical particle-based simulations of systems with unprecedented size and complexity, at experimentally relevant conditions. The framework also enables highly efficient analysis of the resulting simulation data.

MSMs are a powerful approach to simulate dynamics on long timescales;  by performing short simulations to estimate transition rates among system configurations, one can construct an MSM that accurately describes dynamics on timescales that are orders of magnitude longer than the individual simulations \cite{Noe2009, Bowman2009, Pande2010, Sarich2010, Bowman2011, Prinz2011, Voelz2011, Voelz2012, Noe2013, DeSancho2013, Bowman2014, Chodera2014, Malmstrom2014, Schwantes2014, Hummer2015, Husic2018, Zeng2018, Wu2020, Weng2020, Suarez2021}. 
MSMs also provide a means to coarse-grain complex dynamical processes into reduced-order forms that facilitate identifying key slow degrees of freedom and corresponding mechanisms. Furthermore, MSMs can enable designing non-equilibrium assembly protocols that can accelerate assembly and increase selectivity of a specific target state by orders of magnitude in comparison to equilibrium processes \cite{Juarez2012, Tang2016, Tang2017, Grover2019, Trubiano2022}. 

In contrast to previous MSM approaches to self-assembly that pre-assume the state space and transition rates \cite{Jamalyaria2005, Keef2006, Hemberg2006, Dykeman2013, Sweeney2008, Zhang2006, Misra2008, Kumar2010, Xie2012, Smith2014}, we seek a framework in which the state space and transition rates are computed directly from dynamical simulations, and the accuracy of the resulting MSM (including the validity of the Markov assumption) can be directly tested against microscopic dynamics.  While several approaches have been developed to construct MSMs from particle-based assembly simulations \cite{Kelley2008, Perkett2014, Dibak2018, Olsson2017, Sengupta2019, Trubiano2021, Trubiano2022}, these algorithms are designed to track individual assembling clusters evolving under constant conditions such as the concentration of free subunits. Thus, they cannot describe a typical experiment in which a fixed total number of subunits assemble into many structures. In this case, the concentrations of intermediates and free subunits, and thus the transition rates, continuously evolve over time. Moreover, some transitions involve association or dissociation of oligomers or larger intermediates. Therefore, although sophisticated approaches have been recently developed to design optimal assemblies and compute free energy landscapes \cite{Miskin2016, Pietrucci2017, Sidky2018, Sherman2020, Das2021, Henin2022, Wang2022, Das2023, Goodrich2021, Lieu2022, Jhaveri2023, Curatolo2023}, to our knowledge, there is no existing enhanced sampling method that can comprehensively model such self-assembly experiments. 

In this article we present the MultiMSM approach, which provides a complete description of assembly reactions, accounting for changes in transition rates as concentrations evolve, as well as association between intermediates. Using two model systems, we show that that the MultiMSM algorithm accurately predicts the depletion of monomers and the ensemble of resulting intermediate and target assembly species, on the long timescales required for reactions to reach equilibrium. While brute-force dynamics simulations with related models have been limited to restricted parameter ranges, such as high subunit concentrations \cite{Hagan2006,Perlmutter2014,Rapaport2018,Baschek2012,Gupta2023}, the MultiMSM approach enables simulation over a broad range of experimentally relevant parameter values. In particular, the algorithm reduces simulation times by orders of magnitude for systems with large nucleation barriers, which are typically required for productive assembly at experimental conditions \cite{Zlotnick2003, Ceres2002, Gartner2020, Schlicksup2020, Hagan2021, Hagan2023}. 

Crucially, once the MultiMSM has been constructed for one value of the total subunit concentration, assembly dynamics can be simulated over a wide range of lower concentrations without any additional sampling. This enables representing a typical experiment in which assembly is performed over a range of subunit concentrations, but with the computational cost of a single subunit concentration.
Further, 
the method provides a detailed analysis of assembly mechanisms by computing quantities such as the free energy landscape, nucleation timescales, committor probabilities and flux along different assembly pathways, and entropy production rates. The latter quantify the extent to which a reaction is out of equilibrium and identify sources of kinetic trapping that impede productive assembly. These capabilities allow analyzing data from particle-based assembly simulations in unprecedented ways. 

We provide an open-source Python library \cite{Trubiano_SAASH_2024, Trubiano_MultiMSM_2024} which performs all calculations required to construct MultiMSMs and the analysis described in this work, from simulations performed with the open-source molecular dynamics simulation package HOOMD-blue \cite{Anderson2020}. The library can be readily generalized to other software packages.

\begin{figure*}[ht!] 
\centering
    \includegraphics[width=0.98\textwidth]{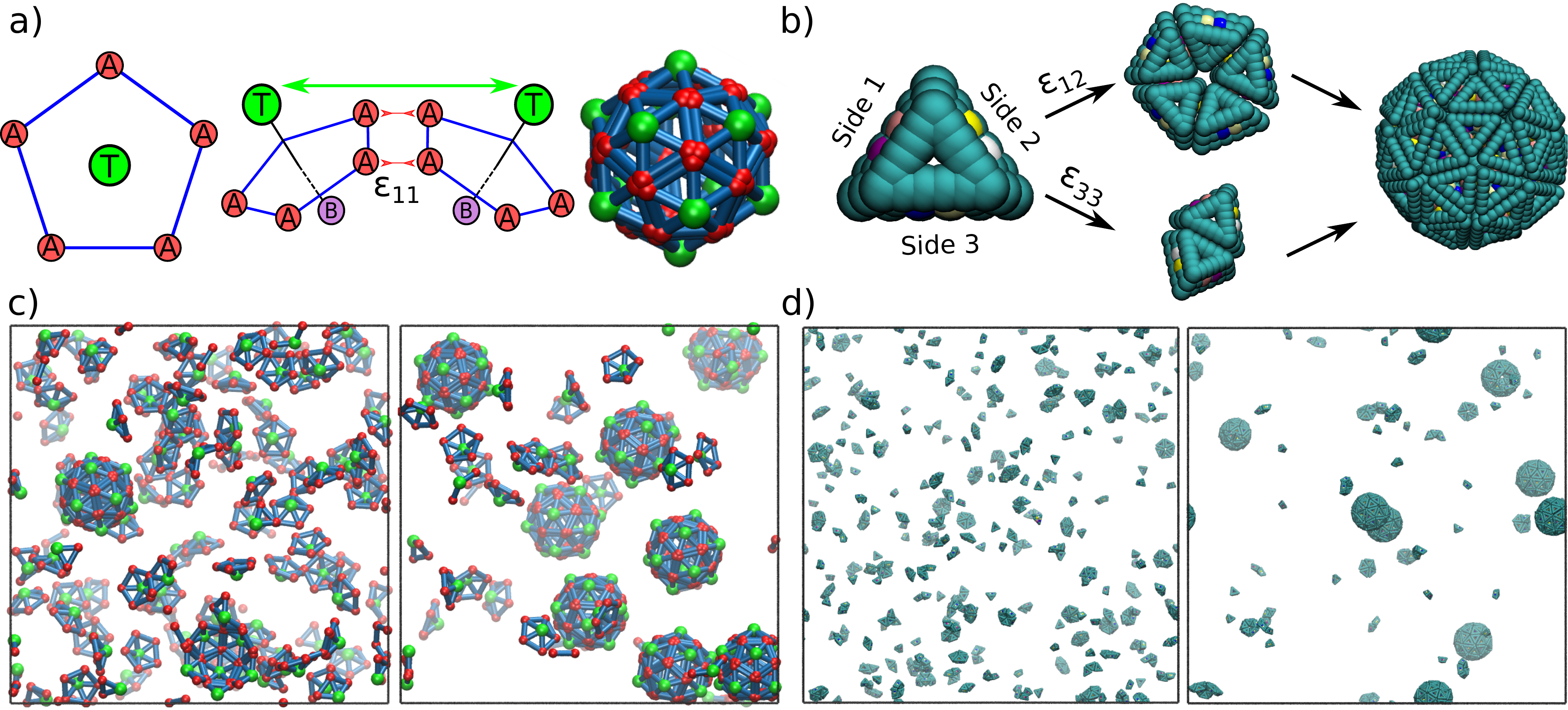}
   \caption{Schematics of the subunits and their interactions for the two model self-assembly systems. 
    \textbf{(a)} (Left) The subunit is a rigid body with Attractors at the vertices of a pentagon and  a `Top' and `Bottom' pseudoatom above and below the vertex plane. (Middle) Subunit-subunit interactions. Attractor-Attractor interactions drive subunit association, with binding affinity parameter $\Ebind$. Top-Top repulsions result in a preferred subunit-subunit binding angle of $116.57^\circ$, making a dodecahedral capsid the ground state. (Right) A snapshot of an assembled capsid from a simulation. 
    \textbf{(b)} (Left) The subunit is a rigid triangular body, motivated by recent DNA origami experiments \cite{Sigl2021,Wei2024}.  The cyan pseudoatoms enforce excluded volume, and complementary pairs of attractor beads (other colors) on each side drive association between Side $1$ -- Side $2$ and Side $3$ -- Side $3$ pairs. (Middle) These interactions, with binding affinity parameters $\Eone$ and $\Ethree$, respectively stabilize pentamers and dimers. (Right) A snapshot of a complete $T=3$ capsid from a simulation. 
    \textbf{(c)} Snapshots of the simulation box during dodecahedron assembly with $\Ebind=5.5$ at early times (left) and late times (right).
    \textbf{(d)} Same as (c) for $T=3$ capsid assembly with $\Eone=11$ and $\Ethree=8$. 
    }
   \label{fig::assembly}
\end{figure*}

\section{Model Systems} \label{sec::models}

We first describe the two model self-assembly systems that we use to test and demonstrate our MultiMSM approach. Our first example, dodecahedral capsid assembly from pentagonal subunits, is sufficiently tractable that brute-force dynamics simulations can be performed on relatively long-time scales to stringently test the MultiMSM results. Our second example, $T=3$ capsid assembly from triangular subunits, is more complex and computationally expensive, and shows that the MultiMSM method scales well for complicated problems with a large state space. 

For all models and results presented in this work, we give energies in units of the thermal energy $\kt$ and lengths and concentrations in units of $\lzero$ and $\lzero^{-3}$ respectively, where $\lzero$ is related to the subunit size for each model (see supplement Section \ref{SMsec::sim_details} \cite{supp}).

\subsection{Dodecahedron Assembly}
\label{sec::dodec}

Our model subunit (Fig.~\ref{fig::assembly}a) is adapted from previous studies of dodecahedral capsids, including assembly of empty capsids and assembly around RNA and synthetic polyelectrolytes \cite{Wales2005, Fejer2009, Johnston2010, Perlmutter2013, Perlmutter2014}. The subunit is a rigid body consisting of five attractor sites, placed at the vertices of a regular pentagon, that have attractive interactions through a Morse potential with well-depth $\Ebind$. Each subunit also has a `Top' and `Bottom' pseudoatom; Top-Top and Top-Bottom pairs on nearby subunits each have repulsive Weeks-Chandler-Anderson (WCA) interactions \cite{Weeks1971}. Top-Top interactions drive subunit-subunit binding angles consistent with a dodecahedron ($116.57^\circ$), while Top-Bottom interactions suppress subunit-subunit binding in inverted orientations \cite{Wales2005, Perlmutter2013, Perlmutter2014}.
We perform simulations in three distinct assembly regimes by setting $\Ebind \in \{5.0, 5.5, 6.0\}$ with total subunit concentration $\ctot=0.0156$, which spans from almost no assembly to rapid assembly. 

\subsection{T=3 Capsid Assembly}
\label{sec::triangles}

Our second example model was previously developed as a simplified representation of an experimental system in which DNA origami forms rigid triangular subunits that assemble into $T=3$ icosahedral capsids \cite{Sigl2021, Wei2024}. The model builds on extensive previous simulations of capsid assembly \cite{Rapaport1999, Rapaport2010, Rapaport2018, Nguyen2007, Nguyen2008, Elrad2010, Williamson2011, Perlmutter2013, Zhang2013, Zhang2014,Zandi2020,Baschek2012,Castelnovo2014,Castelnovo2013,Boettcher2015,Mendoza2020}.   The model subunit excluded volume shape is represented by three layers of `excluder' atoms arranged so that the edges have a bevel angle consistent with an icosahedron ($156.72^\circ$). Each excluder interacts with all pseudoatoms through a WCA potential. Subunit-subunit attractions in the experimental system are driven by DNA blunt-end stacking and hybridization of single-stranded DNA molecules on complementary subunit edges. In the computational model, these short-ranged interactions are represented by placing two `attractor' atoms on each subunit edge, on the middle layer of excluders. Complementary pairs of attractors interact through a Lennard-Jones potential. To match the experimental $T=3$ system, attractors on Side 1 and Side 2 of two interacting subunits are complementary with binding energy (Lennard-Jones potential well-depth) $\Eone$, and attractors on Side 3 are complementary with binding energy $\Ethree$. Attractors that are not complementary interact through a repulsive WCA potential.  Fig.~\ref{fig::assembly}b shows a representation of the triangular subunit, the preferred intermediate for each interaction, and an example of the fully assembled capsid in simulation. Despite the simplicity of the model, Wei et al. 2024 \cite{Wei2024} found that the simulation results semi-quantitatively match experimental observations of capsid assembly dynamics.

Simulations and experiments in Wei 2024 \cite{Wei2024} showed that sufficiently imbalanced values of $\Eone$ and $\Ethree$ lead to hierarchical assembly pathways, since $\Eone$ and $\Ethree$ respectively stabilize intra-pentamer and intra-dimer interactions. Stronger $\Eone$ leads to pentamer-biased assembly pathways, in which subunits rapidly form pentamers, which in turn undergo assembly into capsids; whereas stronger $\Ethree$ leads to dimer-biased pathways, with rapid formation of dimers and their subsequent assembly of capsids. In this work, we focus on parameters that lead to pentamer-biased assembly pathways, $\Eone = 11$ and $\Ethree = 8$, with total subunit concentration $\ctot = 1.7 \times 10^{-4}$. 

All dynamics simulations described in this work were performed with HOOMD-blue \cite{Anderson2020}. Full simulation details can be found in supplement Section \ref{SMsec::sim_details}.

\section{Methods}

Here we describe the protocol to construct and use the MultiMSM. The procedure is separated into four steps: selection of discrete states, data processing and transition counting, monomer fraction discretization and transition matrix construction, and model evaluation and prediction. 

Our python libraries to construct MultiMSMs from HOOMD simulations and perform all the calculations described below are available on Github \cite{Trubiano_SAASH_2024, Trubiano_MultiMSM_2024}. 

\subsection{Selection of Discrete States}
\label{sec::discretization}
We first need to define a state-space discretization, which maps system configurations into MSM microstates. In this section we focus on characterizing the state of an individual assemblage (cluster); we address multiple clusters in section~\ref{sec::TransitionCounting}. We have previously shown that a general state decomposition for assembly is enabled by mapping an assemblage to an undirected graph, with nodes and edges respectively corresponding to subunits and `bonds' (subunit-subunit interactions) \cite{Perkett2014, Trubiano2021}. Alternative approaches based on pairwise distances between subunits and other structural properties have also been used \cite{Karpen1993, Degroot2001, Singhal2004, Andrec2005}. However, these description can be simplified, and the size of the state space significantly decreased, with a simplified state definition that characterizes the number of subunits and bonds within an assemblage \cite{Perkett2014,Perlmutter2014,Trubiano2022}. We use the latter approach for both examples in this article; we define a state as $\state=(\nsub, \bonds)$ where $\nsub$ is the number of subunits in the configuration and $\bonds$ is a count of the number of each type of bond present in the configuration. 
For the examples we consider in this work, $\bonds$ is a scalar for the dodecahedron assembly since there is only one type of bond, but is a vector with two components for the $T=3$ capsid assembly, since there are two interaction types (the Side1-Side2 bond and the Side3-Side3 bond, see Fig.~\ref{fig::assembly}b). We find that these coordinates are sufficient to accurately characterize the dynamics of both systems studied here. 

Since a bond refers to a pair of sufficiently strongly interacting subunits, it must be defined based on a threshold. In this work, we use cutoff distances between corresponding particle types to define a bond. See supplement Section \ref{SMsec::bond_definition} for details on the bond definition for each system.

\subsection{Processing Simulation Data and Counting Transitions}
\label{sec::TransitionCounting}
We seek to model self-assembly dynamics in the canonical (NVT) ensemble. Since there may be many clusters undergoing different stages of nucleation and growth at the same time, we must compute the time evolution of the joint probability distribution of all cluster types $j$. A complete state decomposition would classify the assembly configuration of every cluster at a given time point. However, for a large system with many clusters the number of such states would be intractable. Therefore, we use the independent Markov decomposition (IMD) method \cite{Hempel2021}, in which each cluster is considered as a quasi-independent local subsystem. However, note that the clusters are not strictly independent since pairs of clusters can merge or split during assembly. In this framework, the state probability distribution at frame $i$ is given by \cite{Hempel2021}
\begin{align}
\vec{p}_{\,i}=\nvecsub{p}{1}{i} \otimes  \nvecsub{p}{2}{i} \otimes \ldots \nvecsub{p}{{n_\text{types}}}{i}
\label{eq::pt}
\end{align}
where $\nvecsub{p}{j}{i}$ is the probability distribution for the state of cluster $j$ at frame $i$, $n_\text{types}$ is the total number of cluster-types, and $\otimes$ is the Kronecker product \cite{Hempel2021,Satake1975}. Here, each cluster is defined according to the state decomposition described in section~\ref{sec::discretization}.

For self-assembly, it is useful to cast the cluster probabilities as concentrations
\begin{align}
c^j_i = {\nsub}^j {\nclus}^j_i / V
\label{eq::cj}
\end{align}
where ${\nsub}^j$ is the number of subunits in cluster-type $j$, ${\nclus}^j_i$ is the number of such clusters at a given frame, and  $V$ is the volume. It is important that we use mass-weighted concentrations to maintain the constant total subunit concentration
\begin{align}
\ctot = \sum_j {\nsub}^j c^j_i \quad \forall i
\label{eq::ctot}.
\end{align}
Using number-weighted concentrations would result in a probability that is normalized to the cluster distribution, which depends on time and interaction parameters, whereas the mass-weighted concentrations maintain a normalization that depends only on the control parameter $\ctot$. 

The transition matrix elements can be estimated from the ensemble of short simulations by recording the number of each cluster type $j$ at each simulation window, and then computing the number of transitions between all cluster types, ranging from monomer to dimer transitions to association/dissociation of larger intermediates and complete capsids, as a function of lag time. Importantly, to maintain the mass-weighted cluster distribution, the transition counts need to be weighted by the number of subunits involved in each transition. For example, if a cluster $j$ with ${\nsub}^j$ subunits transitions to a cluster $l$ with ${\nsub}^l>{\nsub}^j$ subunits, then all ${\nsub}^j$ of those subunits undergo the transition. That is, the transition $j\rightarrow k$ occurs ${\nsub}^j$ times. Intuitively, this can be thought of as viewing transitions from the perspective of individual subunits rather than clusters (see supplement Fig.~\ref{SMfig::counting}). 

\textit{Procedure for counting transitions.} For each frame in the simulation, we group the subunits into clusters, with a cluster defined as any collection of more than one subunit that is bonded together. On the first frame, all clusters are given an ID. On subsequent frames, we check if any newly found clusters are derived from existing clusters, either through merging or splitting of sub-clusters, and update any matching existing cluster with the new configuration. In the case of splitting, we record what the parent cluster was and form a new cluster. Any cluster that was not derived from an existing one is assigned a new ID. 

We treat monomers separately, tracking their addition and removal as a separate time series. For example, if an $10$-mer loses two subunits, but they do not form a dimer after dissociating, we log a $-2$ in the monomer time series. If they do form a dimer, we log nothing in the monomer time series, since no monomers are involved in this transition, but record a transition to the dimer. To track monomer-to-monomer transitions, we store the IDs of all subunits that are not bonded in frame $i$ in a list $M_i$. To determine the number of monomer-to-monomer transitions after a lag time $k$, we take the cardinality of the intersection of these lists, $\abs{M_i \cap M_{i+k}}$. 
Finally, we also save the monomer fraction, $\mfrac (i) = \abs{M_i}/\Ntot$, where $\Ntot$ is the total number of subunits, at every frame, and augment any transitions that occur in that frame with this value. 

We then construct the transition count matrix as follows. We loop over every cluster's time series of configurations and extract every pair of configurations separated by a lag time $k$, incrementing the corresponding entry of the count matrix by $1$. Then, to ensure mass-weighting, we multiply that transition count by $\min({\nsub}_{t},{\nsub}_{t+k})$.

\subsection{Monomer Fraction Discretization and Construction of MSMs}
\label{sec::monomerFraction}
In this section we describe how to construct the component MSMs at the different monomer concentrations which arise as subunits are depleted during an assembly reaction. The component MSMs are then combined into the MultiMSM. 

The first step is to discretize the monomer fraction in the interval $[0,1]$, where the monomer fraction is defined as $\mfrac = c^1 / \ctot$ with $c^1$ the concentration of monomers (free subunits).
 The discretization contains $\discIndex+1$ intervals, $D_\discIndex = (0, d_1, d_2, \cdots, d_{\discIndex}, 1)$. We construct an independent MSM on each of these intervals, following the approach described in Section~\ref{sec::TransitionCounting}. 
Although we will describe a smoothing procedure in Section~\ref{sec::smoothing} to interpolate transition matrices between interval edges, the accuracy of the MultiMSM depends strongly on the choice of discretization. While increasing the resolution leads to higher accuracy, it also increases the amount of total sampling required for convergence of the MSMs. Thus, the size of each interval must be chosen such that a sufficiently large number of relevant transitions are sampled. Sampling efficiency can be improved using adaptive sampling \cite{Bowman2010, Doerr2014, Voelz2014, Zimmerman2015, Shamsi2017, Shamsi2018, Zimmerman2018, Wan2020}. 

We have employed a heuristic, but adaptive discretization procedure, in which intervals are defined to contain monomer fraction values that give rise to similar dynamics. For example, in most cases monomer fractions between $0.95$ and $1$ correspond to the initial stage of a reaction when most transitions correspond to monomer-dimer association or dissociation. In contrast, when the monomer fraction approaches its infinite-time limit (e.g. the equilibrium monomer concentration for reversible assembly), most transitions will involve large intermediates.  Ensuring that each interval separates different types of dynamics improves the accuracy of the MultiMSM. We describe a systematic method for refining the monomer fraction discretization in supplement Sections \ref{SMsec::discretization_opt} and \ref{SMsec::error_refinement}. 

We compute the MSM in each interval as follows. We initialize a sparse count matrix to store the number of observed transitions between each pair of states on each of these intervals. Using the output of our cluster analysis at a given lag time (see section~\ref{sec::TransitionCounting}), we identify the monomer fraction at which each transition occurred and increment the count for the corresponding transition and monomer fraction interval. Once all transitions have been logged, we compute the probability transition matrix for each interval by normalizing the rows of the count matrices to sum to $1$.


\subsection{Solving the Forward Kolmogorov Equation}
\label{sec::forward}

The forward equation for the MultiMSM is a straightforward extension of the forward equation for a standard MSM, which is given by 
\begin{equation} \label{eq::FKE}
    \niceVec{p}{n+1} = \niceVec{p}{n} \mat{P}, 
\end{equation}
with $\mat{P}$ as the transition matrix and $\niceVec{p}{n}$ as a row vector giving the probability distribution over discrete states at timepoint $n$. 

For the MultiMSM, the equation has the same form, but involves a collection of transition matrices. At each timepoint $n$ we use the transition matrix corresponding to the current monomer fraction. By construction, we store the monomer fraction in the first component of the probability distribution, $p_0^n$. Formally, we can write
\begin{equation} \label{eq::FKE_multi}
    \niceVec{p}{n+1} = \niceVec{p}{n} \mat{P_{m^n}}, \quad m^n = \text{index}(p_0^n),
\end{equation}
where the index operator converts a value in $[0,1]$ to its corresponding interval in the discretization. 

Note that some of the methods typically used to efficiently solve the forward equation, such as computing the spectral decomposition or pre-computing large powers of the transition matrix, cannot be directly used here since we do not know a priori at what timepoints the monomer fraction will cross the discretization boundaries and change the transition matrix. The most straightforward approach is to solve  Eq.~\eqref{eq::FKE_multi} iteratively via vector-matrix multiplication for each timepoint. However, it is also possible to pre-assume the monomer concentration as a function of time, and then iteratively apply an efficient approach such as spectral decomposition, in which the computed monomer concentration is updated at each iteration.

\subsubsection{Smoothing Solutions}
\label{sec::smoothing}
Solving Eq.~\eqref{eq::FKE_multi} following the above approach typically leads to solutions that are well behaved within each interval of the discretization, but have `jumps' (abrupt changes in slope)  at time steps when a discretization boundary is crossed. These jumps reflect the abrupt change in the transition dynamics due to changing the transition matrix. The jumps introduce small errors in the solution that can accumulate and reduce the accuracy of the MultiMSM prediction at long times. To solve this problem, we describe a smoothing procedure to continuously interpolate between the two transition matrices across a discretization boundary. 

Consider the two intervals closest to $1$, $I_1 = (d_{\discIndex-1}, d_\discIndex)$ and $I_2 = (d_\discIndex,1)$. Let $L_1 = d_\discIndex-d_{\discIndex-1}$ and $L_2 = 1-d_\discIndex$ be the length of each interval and let $\mat{P}_1$ and $\mat{P}_2$ be the transition matrices on intervals 1 and 2, respectively. Since neighboring intervals can have significantly different lengths, we define a smoothing region that is agnostic of absolute interval sizes. Let $\chi$ be the fraction of each interval that is used to smoothly interpolate between them. The transition region will then begin at $a=d_\discIndex - \chi L_1$ and end at $b = d_\discIndex + \chi L_2$. If the monomer fraction falls within $[a,b]$, we construct a linear combination of each interval's transition matrix to use at that value. We choose the weights proportional to where in the region the monomer fraction falls, with an even split if the monomer fraction is precisely $d_\discIndex$. In general, we compute
\begin{equation}
    \alpha(\mfrac)=
    \begin{cases}
        \frac{1}{2} \frac{\mfrac-a}{d_\discIndex-a} & \text{if } a \leq \mfrac \leq d_\discIndex,\\
        \frac{1}{2} + \frac{1}{2} \frac{\mfrac -d_\discIndex}{b-d_\discIndex} & \text{if } d_\discIndex \leq \mfrac \leq b,
    \end{cases}
\end{equation}
where $\mfrac$ is the current monomer fraction. We then construct the final transition matrix as
\begin{equation} \label{eq::LC}
    \mat{P}_m = (1-\alpha(\mfrac)) \mat{P}_1 + \alpha(\mfrac) \mat{P}_2. 
\end{equation}
Since each individual transition matrix is normalized and their coefficients sum to $1$, $\mat{P}_m$ is also normalized and thus a valid transition matrix. 

This smoothing procedure works remarkably well for $\chi \in [0.2,0.3]$. We include $\chi$ as an optional parameter to our solvers, with a default value of $0.25$. Setting a value of $0$ turns off all smoothing and solves Eq.~\eqref{eq::FKE_multi} as stated.

\section{Results and testing of the MultiMSM}

We constructed a MultiMSM using simulation data for each of the model systems described in Section \ref{sec::models}. For each example, we solved Eq.~\eqref{eq::FKE_multi} to predict the time-dependent yields of each discrete state. See Section \ref{SMsec::parameters} in the supplement for a detailed description of how many trajectories were used to build each MSM, adaptive sampling strategies, and the values used for all parameters to the models, such as the monomer fraction discretization.

\subsection{Dodecahedron Capsids}

\begin{figure*}[ht!] 
    \centering
    \includegraphics[width=0.99\textwidth]{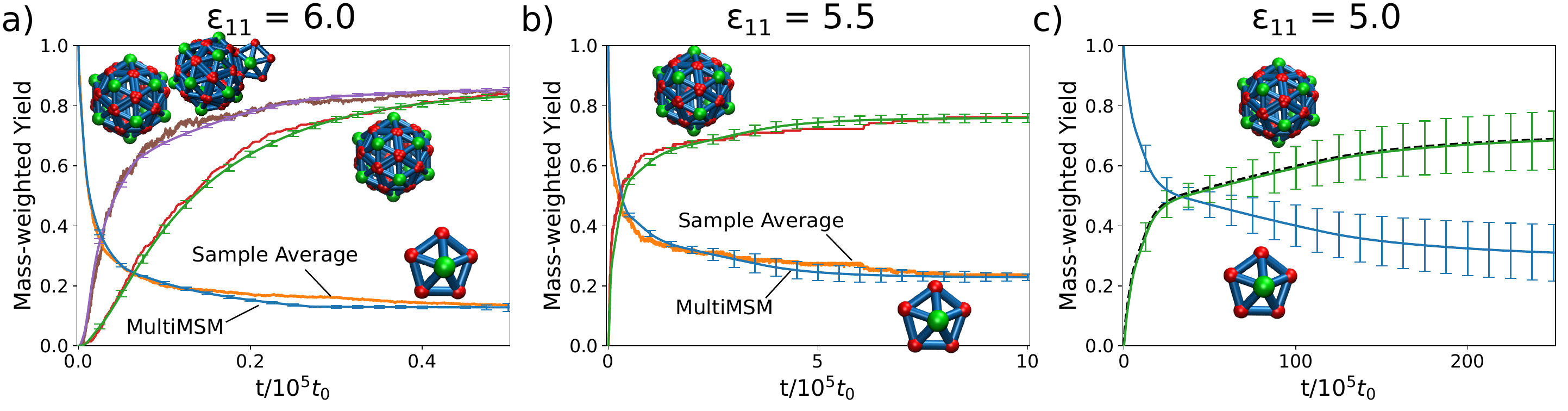}
    \caption{MultiMSM predictions for dodecahedron capsid assembly dynamics at three values of the binding affinity parameter $\Ebind$, compared against results from brute-force dynamics on accessible timescales.  Note the increasing timescale on the x-axis as binding affinity decreases. 
    \textbf{(a)} Strong affinity $\Ebind=6$. The blue, green, and purple curves denote the MultiMSM predictions for mass-fraction of monomer, dodecahedron, and all size 12 structures (capsids and `danglers'), respectively, with representative structures from simulations labeling each curve. The noisy curves (orange, red, brown) show the same mass-fractions estimated from $50$ independent brute-force dynamics trajectories. 
    \textbf{(b)} Moderate affinity $\Ebind=5.5$. MultiMSM predictions for the monomer (blue) and dodecahedron (green) mass fractions; danglers do not form at this binding affinity. The noisy curves (orange, red) show the same mass-fractions estimated from $20$ independent brute-force dynamics trajectories. 
    \textbf{(c)} Weaker affinity $\Ebind=5.0$. MultiMSM predictions are shown for monomers and capsids. The dashed line is $1-\mfrac$ with $\mfrac$ the monomer fraction. 
    The total subunit concentration for (a)-(c) is $\ctot=0.0156$. Error bars are estimated for the MultiMSM by bootstrapping with $1000$ resamplings (see Section \ref{sec::error}). In this work, all energies are given in units of the thermal energy $\kt$ and all length scales in units of $\lzero$ (see text).
    }
    \label{fig::dodec_verify}
\end{figure*}

\begin{figure}[h] 
\centering
    \includegraphics[width=0.49\textwidth]{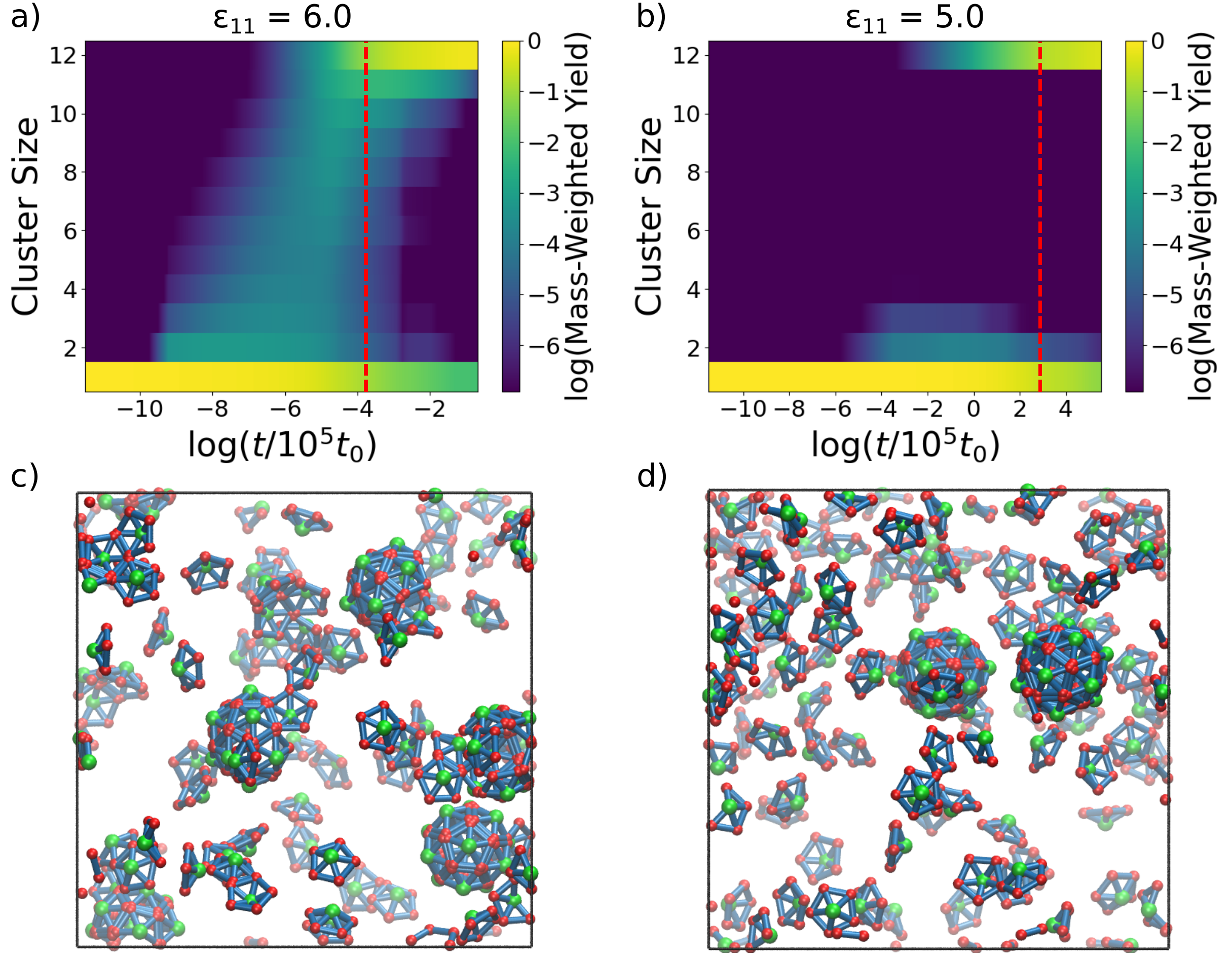}
    \caption{Comparison of MultiMSM intermediate size distributions as a function of time for dodecahedron assembly with \textbf{(a)} $\Ebind=6.0$ and \textbf{(b)} $\Ebind=5.0$. Times and yields are plotted on a log-log scale, over the same time intervals for the corresponding plot in Fig.~\ref{fig::dodec_verify}. \textbf{(c)}, \textbf{(d)} Snapshots of representative configurations  at times corresponding to the red dashed lines in (a), (b). 
    }
    \label{fig::dodec_all_yields}
\end{figure}

Fig.~\ref{fig::dodec_verify} shows example results of assembly dynamics predicted by the MultiMSM for the dodecahedron system at several values of the binding energy $\Ebind$, compared against brute-force dynamics. 
Fig.~\ref{fig::dodec_verify}a shows results for the strongest interactions, $\Ebind=6$, for which assembly is rapid and thus the MultiMSM results can be directly compared against brute-force dynamics simulations on all relevant timescales. Note that for this relatively strong binding energy, it is common for the $12$-th subunit to bind to an $11$-mer in the wrong orientation, and then become trapped for long times. We refer to this off-target, metastable configuration as a `dangler'. The smooth lines (blue, green, purple) in Fig.~\ref{fig::dodec_verify}a show the MultiMSM prediction of the mass-weighted yields of the monomer, dodecahedra, and the size-$12$ structures (dodecahedron and dangler), respectively, while the noisy lines show estimates from brute-force dynamics. The agreement is excellent, with the largest differences being only a few percent at intermediate times, while the short-and long-time behaviors show even closer agreement. 

At early times about half of the $12$-mers are danglers. 
These off-pathway intermediates gradually anneal into the target dodecahedron structure.

Figs.~\ref{fig::dodec_verify}b,c show the MultiMSM predictions for lower binding energy values $\Ebind=5.5$ and $5.0$ respectively. Notably, the weaker subunit-subunit attractions result in significantly longer assembly timescales (about $20\times$ and $500\times$ respectively).  Note that the dangler intermediates do not occur for these weaker binding energies, and thus we focus on the most common structures (monomers and dodecahedrons).
Fig.~\ref{fig::dodec_verify}b shows that the MultiMSM predictions closely match the brute-force dynamics predictions, even on the long timescales required for this system to approach equilibrium. We provide further evidence that this model is accurate over longer timescales in Section \ref{sec::sweeps} and Fig.~\ref{fig::sweeps}b. 
In Fig.~\ref{fig::dodec_verify}c we cannot directly compare the MultiMSM predictions against brute-force dynamics across all timescales, as the simulations would take about $2$ GPU-months per trajectory. We do perform a comparison over accessible timescales ($\TF=2.5\times 10^5 \tzero$) and see good agreement (see supplement Fig.~\ref{SMfig::dodec_E5_verify}). 

   To further test the accuracy of the MultiMSM prediction, we note that the MultiMSM predictions (see Fig.~\ref{fig::dodec_all_yields}), brute-force dynamics simulations, and previous modeling results (e.g. \cite{Zlotnick1994, Endres2002, Hagan2006, Moisant2010}) show that intermediates are present at extremely low concentrations for weak binding energies such as $\Ebind=5.0$. Thus, Fig.~\ref{fig::dodec_verify}c also compares $1-\mfrac$ with the dodecahedron fraction, showing that these two results are within a few percent for all times as expected for small intermediate concentrations. 
Additionally, Fig.~\ref{fig::dodec_all_yields} shows the MultiMSM results for the full cluster-size distribution as a function of time for $\Ebind=6$ and $\Ebind=5$. In the case of stronger binding, there is a broad distribution of detectable intermediate sizes during the rapid assembly phase. In contrast, weak binding results in approximately two-state kinetics --- only dodecahedra and monomer occur at high concentrations, with low concentrations of dimers and trimers, and trace amounts of other transient intermediates. The snapshots in Fig.~\ref{fig::dodec_all_yields}c,d show representative system configurations during the rapid assembly phase for each case. For $\Ebind=6$ (Fig.~\ref{fig::dodec_all_yields}c) we see dodecahedra and monomers coexisting with intermediates of various sizes, while for $\Ebind=5$ (Fig.~\ref{fig::dodec_all_yields}d) we observe two dodecahedra and monomers along with a few transient dimers. 

These results demonstrate a powerful aspect of MSMs that is preserved in our MultiMSM approach; the simulations used to construct the model are all of length $0.2\times 10^5 t_0$, which is orders of magnitude smaller than the relevant assembly timescales.

\subsection{T=3 Capsids}

\begin{figure}[h!] 
\centering
    \includegraphics[width=0.49\textwidth]{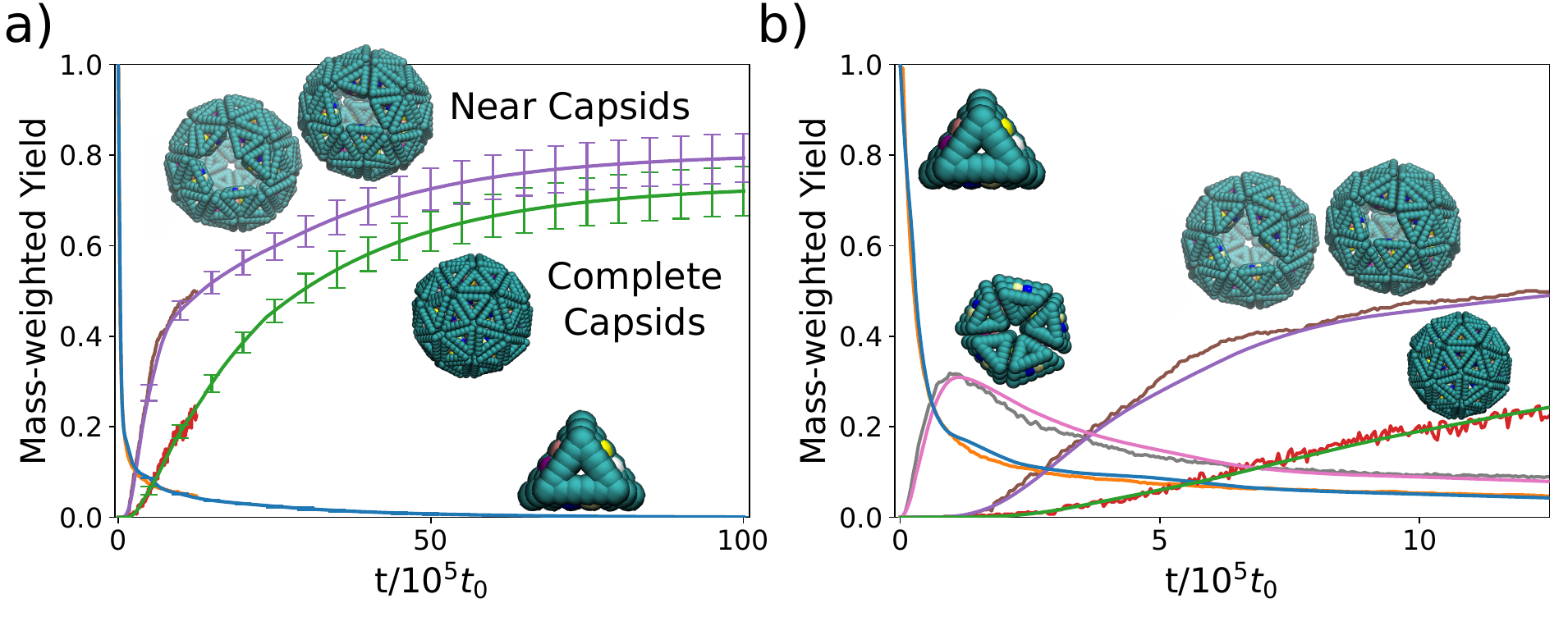}
    \caption{MultiMSM predictions of $T=3$ capsid assembly dynamics with $(\Eone, \Ethree) = (11,8)$ and $\ctot=1.7\times 10^{-4}$, compared against results from brute-force dynamics on accessible timescales. 
    \textbf{(a)} MultiMSM predictions of mass-fractions (smooth curves) are shown for monomers (blue), capsids (green), and `near-capsids' (purple, $56-60$ subunits), with example snapshots labeling each curve. These predictions are compared against mass fractions estimated from $40$ brute-force dynamics trajectories performed up to $12.5\times 10^5 t_0$ (noisy curves). Error bars are estimated for the MultiMSM by bootstrapping with $1000$ resamplings (see Section \ref{sec::error})
    \textbf{(b)} A more detailed comparison of MultiMSM and brute-force results, also showing mass fractions of pentamers (pink). 
    }
    \label{fig::triangle_verify}
\end{figure}

Fig.~\ref{fig::triangle_verify}a shows MultiMSM predictions (smooth curves) for $T=3$ capsid assembly dynamics, with results shown for monomers, complete capsids, and `near-capsids' which include any structure with $56$ or more subunits (including complete capsids with 60 subunits). These predictions are compared against results from brute-force dynamics (noisy curves) on accessible timescales (up to $12.5 \times 10^5t_0$). This comparison is shown in more detail in Fig.~\ref{fig::triangle_verify}b, where we also include pentamers. We observe extremely close agreement, within the statistical error of the brute-force simulations. In particular, the MultiMSM captures the rapid conversion of monomers into pentamers at early times, slow monomer depletion at late times, and the tendency of assembly pathways to become trapped in near-capsid intermediates. 

These behaviors arise because these simulations are performed at binding affinity values $\Eone=11$ and $\Ethree=8$ for which intra-pentamer interactions are strong and intra-dimer interactions are relatively weak \cite{Wei2024}. This imbalance leads to hierarchical assembly pathways in which many subunits first form pentamers, which in turn assemble into nearly complete capsids. However, due to a combination of steric effects, monomer binding, and pentamer depletion, many assembly pathways become trapped in near-capsid structures with $56-59$ subunits. While some of these structures are converted into complete capsids by monomer additions, they persist at $\approx 10\%$ mass fraction even at very long times.


\subsection{Error and Efficiency}
\label{sec::error}


\begin{table}[h!]
\centering
\begin{tabular}{| c | c | c | c | c |} 
 \hline
 & D12 & D12 & D12 & T3  \\ 
 & $\Ebind=6.0$ & $\Ebind=5.5$ & $\Ebind=5.0$ & $(\Eone, \Ethree)=(11,8)$  \\ 
 \hline\hline
 Dynamics & $0.85(6)$ & $0.75(5)$ & $0.11(7)$ & $0.21(8)$ \\ 
 MultiMSM & $0.826(9)$ & $0.74(2)$ & $0.13(3)$ & $0.25(2)$ \\ 
 
 \hline
\end{tabular}
\caption{Comparing the estimated capsid yield and error at $\TF$ for each of the four examples with brute-force dynamics and the MultiMSM. Yields and errors are estimated from brute-force dynamics by sample averages (see Fig.~\ref{fig::dodec_verify} and \ref{fig::triangle_verify}), and from the MultiMSM by bootstrapping with $1000$ resamplings. Final simulation times for the dodecahedron examples are $\TF=0.5\times 10^5 t_0$ for $\Ebind = 6$, $\TF=10\times 10^5 \tzero$  for $\Ebind=5.5$, $\TF=2.5\times 10^5 t_0$ for $\Ebind=5$;  for the $T=3$ capsid $\TF=12.5\times 10^5 t_0$.
 }
\label{table::bootstrapping}
\end{table}


\textit{Error.}
For a standard MSM, the uncertainty in the equilibrium distribution and other quantities can be directly propagated from uncertainty estimates in transition matrix entries \cite{Hinrichs2005, Hinrichs2007, Bhattacharya2019, Kozlowski2023}. However, in the MultiMSM such propagation is complicated by the unknown switching times between the component MSMs and the smoothing procedure. Therefore, we quantify errors using bootstrapping \cite{Efron79, Efron81} (see supplement Section \ref{SMsec::bootstrapping} for further details).  

Table \ref{table::bootstrapping} shows a comparison of the estimated means and standard errors
of the capsid yield for each of the examples from the MultiMSM by bootstrapping with $1000$ resamples. These results are compared against sample averages from the brute-force dynamics simulations. In each example the comparison is shown for the final simulation time point $\TF$. We see that the MultiMSM yields are within the statistical error of the estimates from brute-force dynamics, and that the statistical error for the MultiMSM is consistently smaller than that from the dynamics. 

We have used the same bootstrapping approach to compute the statistical error of the MultiMSM yield predictions as a function of time for each of the assembly examples (Fig.~\ref{fig::dodec_verify} and \ref{fig::triangle_verify}a). 
Where available, the sample-averaged yields are typically within the error bars of the MultiMSM prediction. 
An exception is dodecahedron assembly with $\Ebind=6$ (Fig.~\ref{fig::dodec_verify}a), for which the computed error bars are quite small at some times and sample averages lay outside them. This is likely because this example used the fewest sample trajectories to build the MultiMSM and we did not perform adaptive sampling, so the sampling with replacement step performed for the bootstrap results in very similar models. 
For the examples with longer assembly timescales (Fig.~\ref{fig::dodec_verify}c and \ref{fig::triangle_verify}a), the errors generally grow in time due to accumulated error from each monomer fraction interval of the MultiMSM. While the error for dodecahedron assembly with $\Ebind=5$ is particularly large at long time scales, our analysis shows that this is because we had very limited sampling at the small monomer fraction values that occur at long timescales. Importantly though, this error could be significantly reduced with further sampling.  The error estimates provide a guide to refining the discretization of the monomer fractions and performing additional sampling (see supplement Sections \ref{SMsec::adaptive_sampling} and \ref{SMsec::error_refinement}). 


As for a traditional MSM, the accuracy of the MultiMSM at long times is limited by the sampling of the most relevant slow transitions. For example, in the triangles system which involves strong intra-pentamer interactions, we sampled pentamer-to-monomer transitions only $25$ times and dissociation from complete capsids only $\sim 500$ times. 
As noted above, adaptive sampling techniques focus on such transitions to improve statistics. However, estimates of transition matrix elements that involve such rare events can be improved much more efficiently by incorporating non-Boltzmann techniques \cite{Trendelkamp-Schoer2016, Noe2019, Bonati2021, Falkner2023}.

\begin{figure}[h!] 
\centering
    \includegraphics[width=0.5\textwidth]{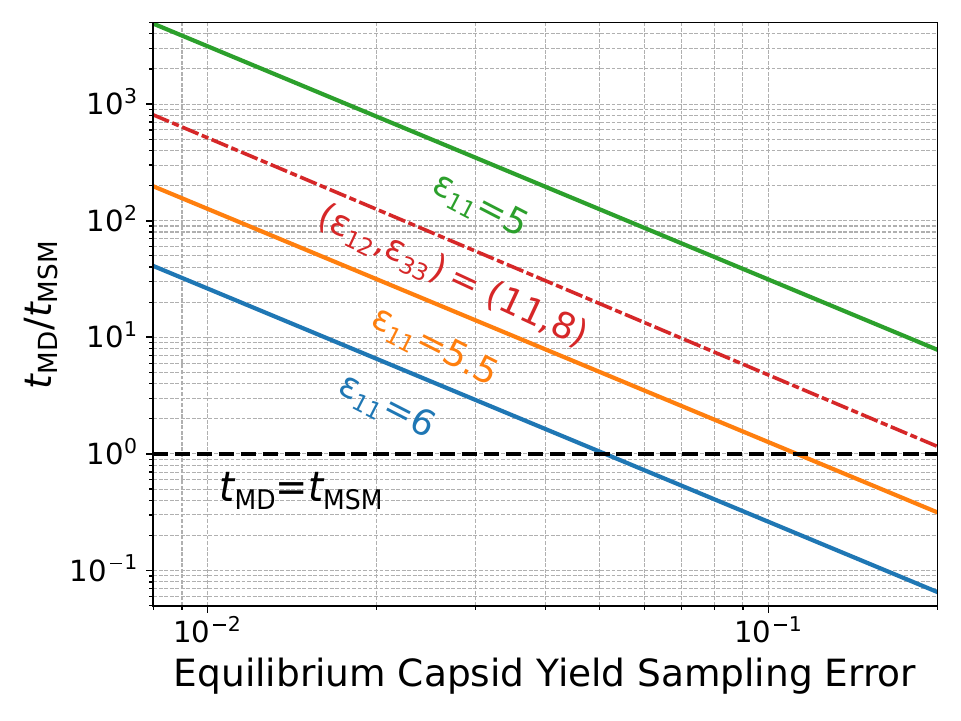}
    \caption{Estimate of the computational speedup provided by the MultiMSM. 
    The plot shows the ratio of the total simulation time for constructing a MultiMSM, $t_\text{MSM}$, and the total simulation time to estimate the equilibrium capsid yield to a given error tolerance by sample averaging across brute-force dynamics trajectories, $t_\text{MD}$, for the four example systems considered in this work: dodecahedron assembly with $\Ebind=6$ (blue), $\Ebind=5.5$ (orange), $\Ebind=5$ (green), and $T=3$ capsid assembly (red) with $\Eone=11$ and $\Ethree=8$. 
    We compute $t_\text{MD}$ as the number of GPU-hours to simulate a system until the capsid mass fraction reaches $99 \%$ of its equilibrium yield (as determined from the MultiMSM results), multiplied by the expected number of independent trajectories, $N$, needed to reach a given sample error tolerance (assumed to scale as $\sigma_0/\sqrt{N}$, where $\sigma_0$ is fit for each example using data in Table \ref{table::bootstrapping}). 
    The dashed line denotes an efficiency ratio of one, above which the MultiMSM is more efficient. }
    \label{fig::efficiency}
\end{figure}


Next we compare the efficiency of constructing a MultiMSM compared to performing an ensemble of straight-forward dynamics simulations, with the estimate of the equilibrium capsid yield as a benchmark. Fig.~\ref{fig::efficiency} shows the \textit{efficiency} as a function of sampling error for each example system, where we have defined the efficiency as the ratio of computational time for brute-force dynamics required to simulate dynamics until the average capsid mass fraction reaches $99\%$ of its equilibrium value with the specified error, relative to the time required to build a converged MultiMSM with the same accuracy.  The time required to perform the bootstrapping with the MultiMSM is negligible compared to the simulation time, so we exclude this time from the calculation. We obtain similar results for other metrics of comparison between the brute-force dynamics and MultiMSM. The horizontal dashed line shows the threshold above which the MultiMSM is more efficient.

We see that for all systems the MultiMSM is significantly more efficient than brute-force dynamics for acceptable error levels, with better than $10^4$ speedup to achieve $1\%$ error for the dodecahedron system with $\Ebind=5$. The speedup is significantly lower for the $T=3$ system --- this is because, to enable direct comparison between the MultiMSM predictions and brute-force dynamics, we have run that system for a strong binding affinity with correspondingly small nucleation barriers and rapid assembly. 
As with any MSM algorithm, the MultiMSM speedup depends on the separation of timescales. For self-assembly, the speedup will increase exponentially with the height of nucleation barriers; i.e., with decreasing binding affinity or subunit concentration.

\section{Applications}

\subsection{Concentration Sweeps}
\label{sec::sweeps}

A powerful consequence of how the MultiMSM is constructed is that total subunit concentrations $\ctot$ below that used in the simulations to build the MultiMSM can be analyzed with no additional sampling. Consider a discretization of the monomer fraction, $D = [0, d_1, ..., d_\discIndex, 1]$, such that a monomer fraction of $1$ corresponds to a maximum total subunit concentration of $\ctotMax$. 
For our primary analysis, we initialize the system using the MSM defined on the interval $[d_\discIndex, 1]$ with a starting distribution of all monomers, $p_i^0 = \delta_{i0}$, and the solution to Eq.~\eqref{eq::FKE_multi} gives the system dynamics with concentration $\ctotMax$.

\begin{figure*}[ht!] 
\centering
    \includegraphics[width=0.98\textwidth]{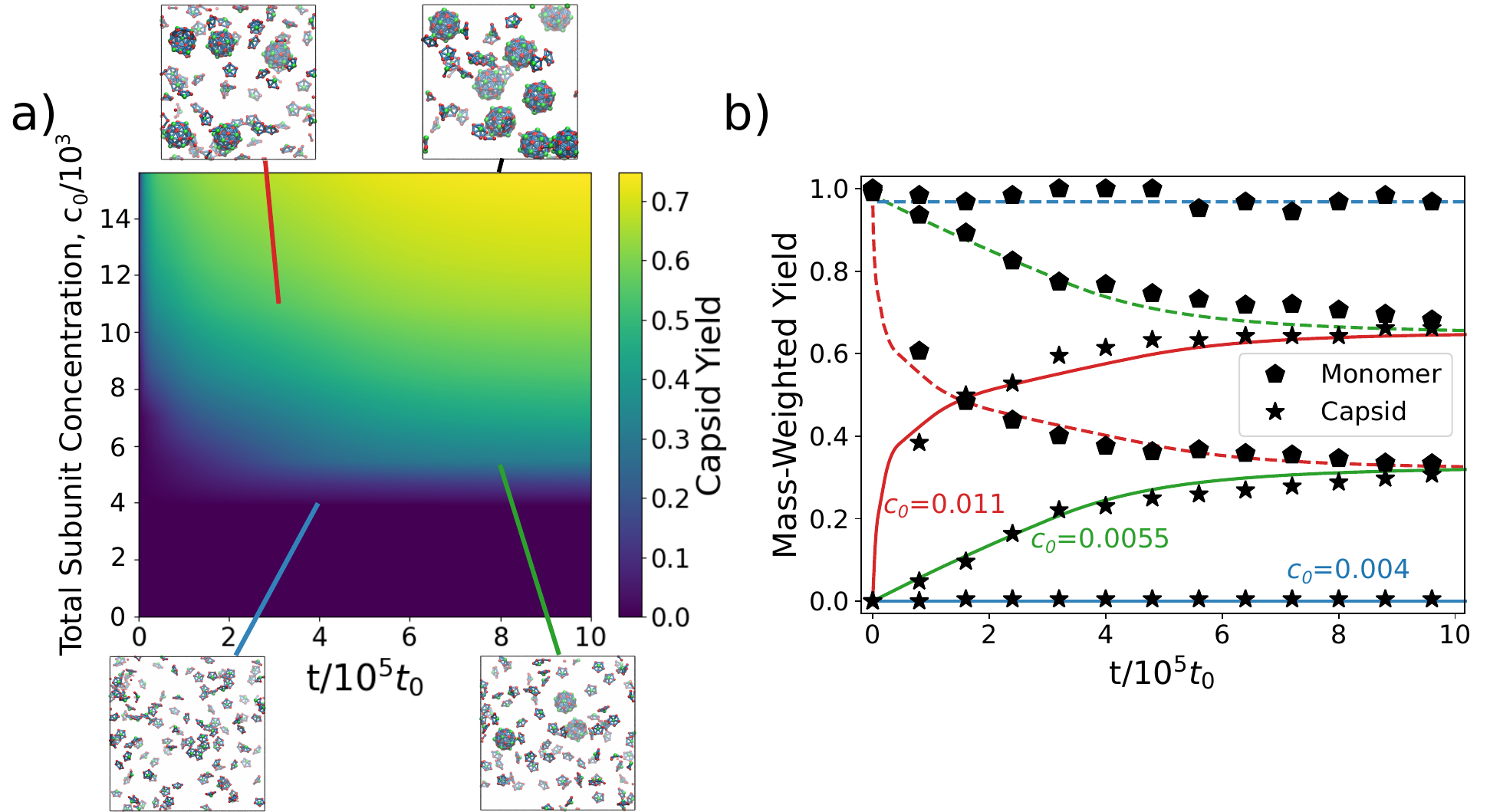}
    \caption{Using the MultiMSM to estimate how assembly dynamics depends on total subunit concentration, $\ctot$, without requiring additional simulations. 
    \textbf{(a)} The MultiMSM prediction for the capsid mass fraction is shown as a function of time and $\ctot$ for the dodecahedron assembly with $\Ebind=5.5$. Representative snapshots of the simulation box are shown for four choices of time and total subunit concentration. 
    \textbf{(b)} Comparison of the MultiMSM predictions of monomer (dashed lines) and capsid (solid lines) mass fractions against brute-force dynamics results (symbols) for total subunit concentrations $0.7\ctotMax$ (red), $0.35\ctotMax$ (green), and $0.25\ctotMax$ (blue). For the brute-force dynamics simulations, the total number of subunits was fixed at $\Ntot=125$, the box size was increased to $\Lzero=22.525\lzero$, $28.380\lzero$, and $31.750\lzero$ respectively, and $20$ independent trajectories were performed for each system. Results from the brute-force dynamics simulations at lower concentrations were not used in construction of the MultiMSM. 
    }
    \label{fig::sweeps}
\end{figure*}

We can then approximate the dynamics for lower subunit concentrations  $\ctot=d_k \ctotMax$ without any additional sampling, by initializing the MSM defined on one of the inner intervals, $[d_{k-1,} d_{k}]$ for $k<\discIndex$, with the same initial distribution of all monomers. We define a new monomer fraction discretization $D_k = [0, d_1/d_k, d_2/d_k, ..., d_{k-1}/d_k, 1]$, and assign corresponding transition matrices to the same intervals. For example, if $\mat{P}_1$ originally was the transition matrix on the interval $[0, d_1]$, in the reduced system it will be the transition matrix on the interval $[0, d_1/d_k]$. 
A limitation is that we are constrained to the finite set of concentrations corresponding to the bins of our discretization, $d_k\ctot$, and depending on the quality and amount of sampling initially performed in the bins closer to zero, the error may increase as smaller concentrations are probed. However, if needed, additional sampling can be performed to refine the discretization based on error analysis as described in Section~\ref{sec::error} and supplement Section \ref{SMsec::error_refinement}.

Fig.~\ref{fig::sweeps}a shows the results of a concentration sweep performed using this method for the dodecahedron system with $\Ebind=5.5$. The capsid yield curves are computed for the initial concentration $\ctotMax$ and smaller values corresponding to the monomer fraction discretization bin cutoffs, and then interpolated between these discrete values. The results show that the final capsid yields decrease while assembly timescales increase as the total subunit concentration is reduced, consistent with previous theory and experiment \cite{Zlotnick1994, Zlotnick1999, Endres2002, Zlotnick2003, Hagan2010, Mizrahi2022, Zlotnick2003a,Mateu2013,Perlmutter2015,Hagan2014,Zandi2020,Prevelige1993,Schwartz1998,Sweeney2008,Zhang2006a,Tan2021,Yu2021,Panahandeh2020,Rapaport2008,Whitelam2009,Nguyen2007,Wilber2007,Wilber2009,Hagan2011,Cheng2012,Rapaport2004,Nguyen2009,Johnston2010,Rapaport2010,Nguyen2008,Bruinsma2015,Hagan2016,Rapaport2012,Rapaport1999,Mendoza2020,Wagner2015a,Castelnovo2014,Castelnovo2013,Boettcher2015,Schoot2007}.  We can also infer the critical assembly concentration, as the MultiMSM initialized at concentration $0.25\ctotMax$ predicts a capsid yield of zero.  

This procedure provides an excellent approximation to the dynamics when there is a large separation of timescales between nucleation and growth timescales, so that monomers are depleted slowly in comparison to the timescale for transitions among larger intermediates. Such a separation of timescales is consistent with the usual criteria for MSMs to provide effective computational speed up. The dodecahedron assembly examples with lower binding energies ($\Ebind=5.0, 5.5$) are good examples of this scenario. Cluster nucleation is a rare event, and entire transition pathways from monomer to capsid are frequently sampled in each of the monomer fraction discretization bins. Consequently the approximation is highly accurate in these cases, as shown in  Fig.~\ref{fig::sweeps}b, which compares the MultiMSM predictions made by extrapolating to lower total concentrations against brute-force dynamics trajectories. We performed simulations until $10\times 10^5t_0$, when the capsid yield reaches $\approx 99\%$ of its equilibrium value according to the MultiMSM for $\ctot=0.7\ctotMax=0.011$ and $\ctot=0.35\ctotMax=0.0055$. For $\ctot=0.25\ctotMax=0.004$, the MultiMSM predicts no assembly, and we observe only a single capsid out of a possible $200$ in the brute-force dynamics. These comparisons provide further evidence that the transition matrix estimation is accurate in the lower monomer fraction bins of the original MultiMSM. This reinforces the accuracy of the long-time predictions at the original concentration $\ctotMax$ in Fig.~\ref{fig::dodec_verify}b, which depends on the accuracy of the transition matrix on each interval. 

For the example in Fig.~\ref{fig::dodec_verify}c, approaching equilibrium with brute-force simulations is computationally intractable, and thus we cannot directly test the MultiMSM predictions. Importantly though, performing this concentration sweep out to such timescales using the MultiMSM takes only a few minutes on a CPU, demonstrating many orders of magnitude of computational speed-up.

\begin{figure}[h!] 
\centering
    \includegraphics[width=0.5\textwidth]{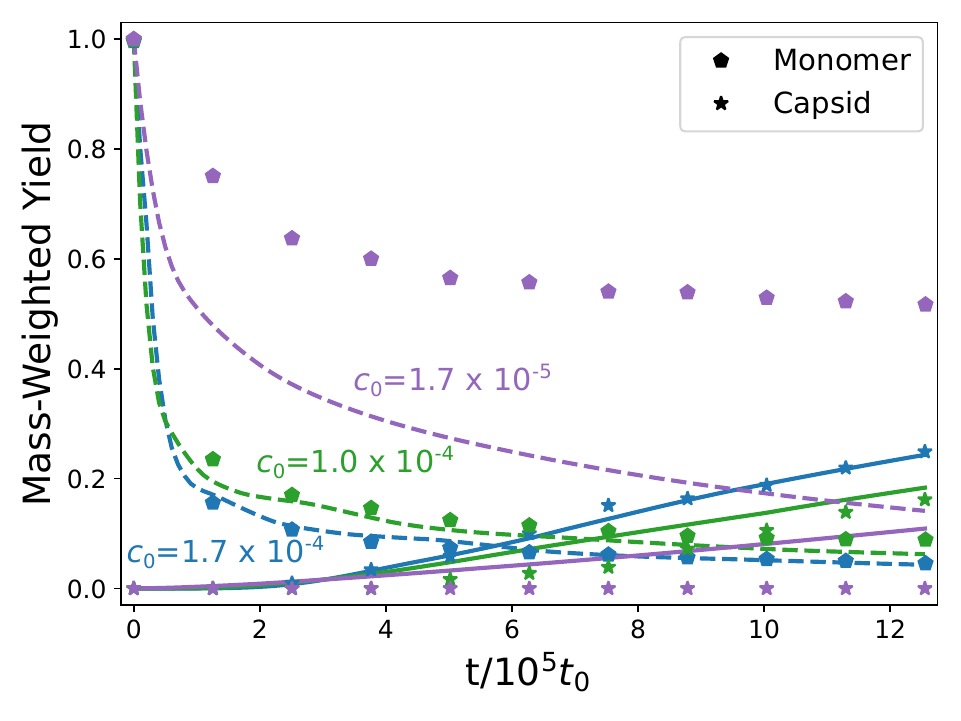}
    \caption{ The $T=3$ capsid assembly example shows that the concentration sweep approximation can become inaccurate for insufficient separation of timescales. Comparison of the MultiMSM predictions of monomer (dashed lines) and capsid (solid lines) mass fractions against brute-force dynamics results (symbols) for total subunit concentrations $\ctotMax$ (blue), $0.63\ctotMax$ (green), and $0.1\ctotMax$ (purple). For the brute-force dynamics simulations, the total number of subunits was fixed at $\Ntot=600$, the box size was set to $\Lzero=153\lzero$, $179.4\lzero$, and $331.4\lzero$ respectively, and $20$ independent trajectories were performed for each system. Results from the brute-force dynamics simulations at lower concentrations were not used in construction of the MultiMSM.  }
    \label{fig::sweep_triangles}
\end{figure}

The approximate concentration results are less accurate in cases with poor separation of timescales. The triangle system at the simulated binding affinity provides an example of this scenario, in which monomers deplete quickly, but larger intermediates form over a longer timescale. 
Fig.~\ref{fig::sweep_triangles} shows the result of performing the concentration sweep for the $T=3$ capsid assembly, comparing to brute-force dynamics simulations over the computationally accessible timescale of $\TF=12.5\times 10^5 \tzero$. The blue curves used the initial total subunit concentration, $\ctot=\ctotMax$, as a reference, while the green and purple curves used concentrations $\ctot=0.63\ctotMax=1.0\times 10^{-4}$ and $\ctot=0.1\ctotMax=1.7 \times 10^{-5}$, respectively. 
We see that for the intermediate concentration (green), $\ctot=1.0\times 10^{-4}$, the MultiMSM predictions match well with the brute-force estimates except for a slight overestimate of capsid yields at early times. Monomer depletion is still accurately characterized by the MultiMSM in this case. 
However, for lower concentrations such as $\ctot=1.7 \times 10^{-5}$ (purple), the method breaks down; the MultiMSM predicts too rapid monomer depletion and a nonzero capsid yield, even though the largest intermediate size is $10$ in brute-force simulations on this timescale. This breakdown can be attributed to the relatively high populations of intermediates; once the monomer fraction becomes small enough ($<10\%$) in the initial system, most of the sampled transitions are between monomers and larger intermediates. When we then use these transition matrices to predict the dynamics of a system with all monomers in the approximate concentration sweep, the prediction overestimates the rate of monomers forming larger intermediates. 
This breakdown of the concentration sweep approximation can be identified by comparing the yield of intermediates to the total subunit concentration. When the triangle monomer fraction reaches $0.63$, intermediates with sizes between $5$ and $55$ subunits account for $\sim 10\%$ of the yield, and the concentration sweep remains accurate when initialized at this total subunit concentration. When the triangle monomer fraction reaches $0.1$, the same set of intermediates account for over $55\%$ of the yield and the concentration sweep approximation breaks down. In our examples, an intermediate yield of about $20\%$ seems to be an upper bound for getting reasonable results from the MultiMSM concentration sweep. 



\subsection{Transition Path Theory}

Transition path theory (TPT) is a powerful tool to study the statistics of reactive trajectories, in which the system transitions between $2$ (meta)stable states \cite{Vanden-Eijnden2006, Metzner2009, Noe2009, Weinan2010}. For self-assembly, TPT provides detailed information about kinetics, assembly pathways, and related quantities. 
TPT is well established for equilibrium systems with stationary dynamics, and has been previously applied to self-assembly systems \cite{Perkett2014}. However, the MultiMSM is inherently non-stationary and non-equilibrium, so we must extend the TPT framework for self-assembly to this context.

We take the approach of Helfmann to generalize TPT to non-stationary, finite-time systems \cite{Helfmann2020}.  
The main quantity of interest in transition path theory is referred to as the committor, $\vec{q}$. Given an initial state (free monomers in our case) and a final state (e.g. a complete capsid in our system), the \emph{forward} committor element $q_i^+(n)$ gives the probability of reaching the target state before returning to the initial state, given the system begins in an intermediate state, $i$, at time $n$. If we let $A$ denote the set of initial states and $B$ denote the set of target states, the forward committor can be computed as the solution of the iterative equations:
\begin{align} 
    q_i^+(n) &= \sum_j P_{ij}(n)q_j^+(n+1), \quad i \notin A \cup B, \label{eq::fcommittor}\\
    q_i^+(n) &= 0, \quad i \in A,\\
    q_i^+(n) &= 1, \quad i \in B,
\end{align}
with a final condition $q_i^+(N) = 1_B(i)$ where $\vec{1}_B$ is the indicator vector for the set $B$, and $P_{ij}(n)$ is the $(i,j)$ component of the transition matrix at time $n$. This equation can be solved for all time by starting at the final condition and backpropagating the solution to earlier times via the transition matrix, while explicitly enforcing the boundary conditions at every time step. Note that this calculation requires knowing the transition matrix at every time step, so Eq.~\eqref{eq::FKE_multi} must be solved first (see supplement Section \ref{SMsec::backward_eq}).  

In addition to the forward committor, we can also compute a \emph{backward} committor, $q_i^-(n)$, which gives the probability that a trajectory being in state $i$ at time $n$ came most recently from set $A$ rather than $B$. This quantity is governed by the transition matrix of the time-reversed process, which is given by
\begin{equation}
    P_{ij}^-(n) = \frac{p_j^{n-1}}{p_i^n} P_{ji}(n-1),
\end{equation}
where $p_i^n$ is the probability of being in state $i$ at time $n$ according to Eq.~\eqref{eq::FKE_multi}. Note that for a stationary, equilibrium process, this is simply a statement of detailed balance. The backward committor can be computed from a similar set of iterative equations:
\begin{align} 
    q_i^-(n) &= \sum_j P_{ij}^-(n)q_j^-(n-1), \quad i \notin A \cup B, \label{eq::bcommittor}\\
    q_i^-(n) &= 1, \quad i \in A,\\
    q_i^-(n) &= 0, \quad i \in B,
\end{align}
with initial condition $q_i^-(0) = 1_A(i)$ with $\vec{1}_A$ as the indicator vector for the set $A$. These equations can be solved simply by forward propagation from the initial condition, which can be done concurrently while solving Eq.~\eqref{eq::FKE_multi} for $\niceVec{p}{n}$. After computing both committors, the current or flux of reactive trajectories can be defined as 
\begin{equation}\label{eq::flux}
    f_{ij}^{AB}(n) = q_i^-(n) p_i^n P_{ij}(n) q_j^+(n+1),
\end{equation}
with effective current $f_{ij}^+(n) = \max\{f_{ij}^{AB}(n) - f_{ji}^{AB}(n), 0\}$. 

These TPT calculations are well-suited to the MultiMSM framework; Eqs.~\eqref{eq::FKE_multi} and \eqref{eq::bcommittor} can be solved in the forward pass of the model, and then Eqs.~\eqref{eq::fcommittor} and \eqref{eq::flux} can be solved with a backward pass. However, we must consider mass-weighted transitions in the transition matrix, and focus the committer calculation on the largest cluster involved in the transition. For example, when an $n$-mer loses a subunit, we log a transition to both the $(n-1)$-mer and the monomer. The committor would be greatly underestimated if all intermediate states were able to transition directly back to the initial state. We describe how to account for these complications in supplement Section \ref{SMsec::clusterization}. 

\begin{figure}[ht!] 
\centering
    \includegraphics[width=0.49\textwidth]{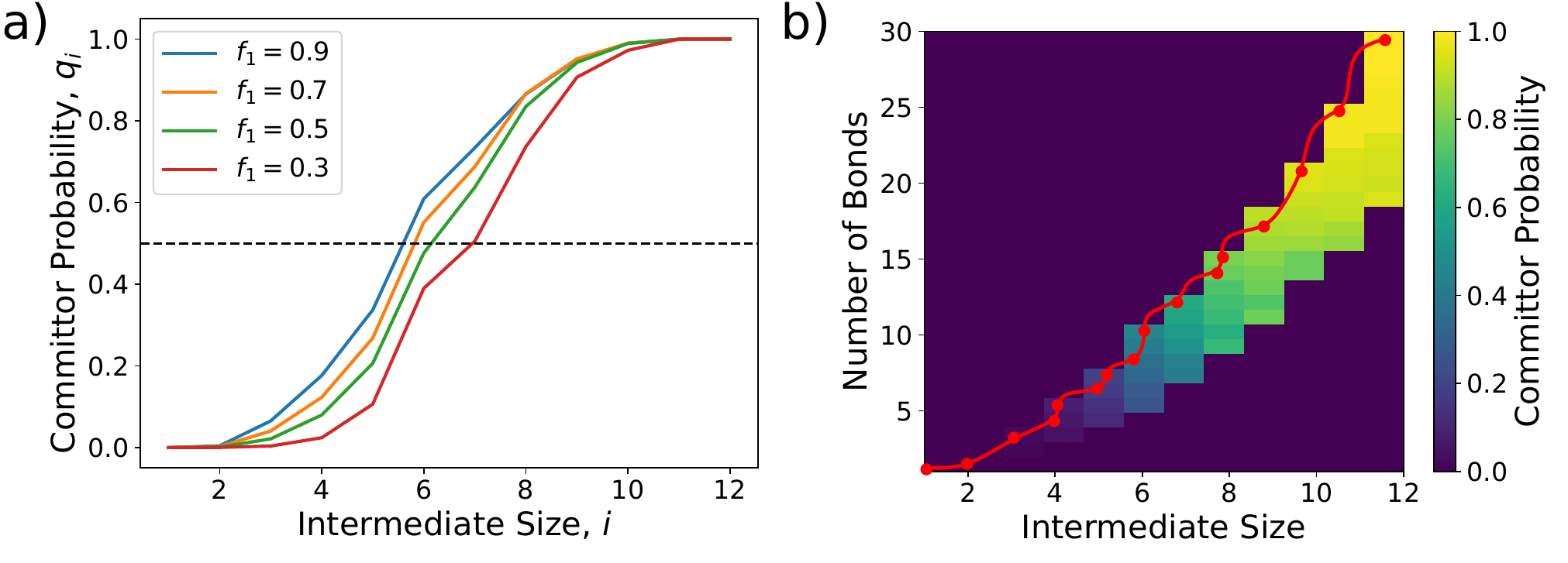}
    \caption{Transition path theory (TPT) on the MultiMSM for dodecahedron capsids with $\Ebind=5.5$ and $\ctot=0.0156$. The initial state is entirely monomers and the target state is a complete capsid with $(\nsub, \bonds) = (12,30)$. 
    \textbf{(a)} The forward committor probability $q_i$ as a function of intermediate size, $i$, for indicated values of the monomer fraction $\mfrac$. Committor values are taken as a maximum over all configurations of a given size at the first time step for which the monomer fraction falls below the reported value. A dashed line is plotted at probability $0.5$; the intersection points with the committor curves denote the critical nucleus sizes. 
    \textbf{(b)} The forward committor probabilities in the full $(\nsub, \bonds)$ state space at the first time the monomer fraction falls below $\mfrac=0.5$. The red, dotted line denotes the most likely reactive path from the initial to target state. 
    }
    \label{fig::tpt}
\end{figure}

Fig.~\ref{fig::tpt} shows the results of performing these TPT calculations for the dodecahedron system with $\Ebind=5.5$.  In this example, the initial state is entirely monomers, $(\nsub, \bonds) = (1,0)$, and the target state is a complete capsid with $(\nsub, \bonds) = (12,30)$.  
Fig.~\ref{fig::tpt}a shows the forward committor probability as a function of intermediate size, with each curve corresponding to a different time point along the course of the reaction, and thus a different free monomer fraction $\mfrac$. The intermediate size at which the forward committor is $50\%$ (dashed line) gives an estimate of the critical nucleus size, which thus can be identified as the intersection between the dashed line and a given committor curve. Importantly, we note that statistical tests would be required to establish whether the intermediate size is a good reaction coordinate, and what, if any, additional features would need to be added to the reaction coordinate to give a rigorous description of the ensemble of critical nucleus structures \cite{Dellago2002, Peters2016}. However, we expect the trend shown in Fig.~\ref{fig::tpt}a to be reliable, and the results can be tested against the free energy calculations shown in Section~\ref{sec::free_energy}.
We find that as monomers are depleted, the committor curves shift toward larger intermediate sizes and the critical nucleus size increases from about $5$ subunits when $\mfrac=0.9$ to $7$ subunits when $\mfrac=0.3$. This trend reflects the fact that while disassembly rates are independent of $\mfrac$, association rates to a growing capsid are proportional to $\mfrac$. Or, from a thermodynamic viewpoint, the chemical potential cost of removing a free monomer from solution grows as $\kt \ln \mfrac$. 

Fig.~\ref{fig::tpt}b shows the forward committor values for the dodecahedron system when $\mfrac=0.5$, but in the full $(\nsub, \bonds)$ state space. We also plot the most probable reaction pathway from a monomer to a capsid (red line and points). 
This path is computed by determining the sequence of states between monomer and capsid with the largest effective flux, given by $f_{ij}^+$. The resulting path matches observations from previous calculations of capsid assembly \cite{Endres2002, Hagan2014, Mizrahi2022}; the most probable assembly pathway proceeds through intermediates with the maximum number of bonds between subunits, or those intermediates that can easily rearrange to form new contacts and thus increase the number of bonds without the addition of new subunits.  We obtain similar results for other monomer fraction values, except that committor probabilities shift to slightly higher intermediate values with decreasing $\mfrac$. The most probable path is not sensitive to the monomer fraction, although sometimes the transitions that involve rearranging the bonds at fixed intermediate size occur in one step rather than two.

\subsection{Entropy Production Rates}

The entropy production rate provides a means to quantitatively measure how far from equilibrium a process is, thus elucidating the irreversibility of a process (which produces entropy), the heat produced or work done by the system, or the efficiency of a process \cite{Bag2002, Vaikuntanathan2009, Dibyendu2017, Landi2021}. 
Much recent work has developed optimal control algorithms for non-equilibrium systems to minimize entropy production  \cite{GomezMarin2008, Sivak2016, Large2019, Tafoya2019, Gingrich2016, Nakazato2021, Dechant2022, Blaber2023, Brown2020, Abiuso2022, Louwerse2022, Whitelam2022, Zhong2022, Gupta2023, Davis2024}. However, entropy production is frequently difficult to measure in experiments or simulations of complex models, due to the large amount of data needed to reliably estimate probability distributions and currents, although approaches based on machine learning \cite{Kim2020, Nir2020, Otsubo2020, Otsubo2022, Boffi2023a, Boffi2023b, Whitelam2023} and automatic-differentiation \cite{Engel2023} can help. 
Fortunately, the MSMs enable computationally efficient computation of entropy production.

The time-dependent entropy production rate for a Markov chain at step $n$ is given by the expression
\begin{equation} \label{eq::entropyMC}
    \ep^n = \frac{1}{2} \sum_{i,j} \left( p_{i}^n P_{ij} - p_j^n P_{ji} \right) \log \frac{p_i^n P_{ij}}{p_j^n P_{ji}},
\end{equation}
where $p_i^n$ is the probability of being in state $i$ at time $n$, and $P_{ij}$ is the transition probability from state $i$ to state $j$. This extends naturally to the MultiMSM framework, with the transition matrix replaced by the time-dependent transition matrix across the monomer fraction bins
\begin{equation} \label{eq::entropyMultiMSM}
    \ep^n = \frac{1}{2} \sum_{i,j} \left( p_{i}^n P_{ij}(n) - p_j^n P_{ji}(n) \right) \log \frac{p_i^n P_{ij}(n)}{p_j^n P_{ji}(n)},
\end{equation}
where $\mat{P}(n)$ is the transition matrix used to update the system at time step $n$. 
This quantity can be computed while solving the forward Kolmogorov equation as described in Section \ref{sec::forward}. 

\begin{figure}[ht!] 
    \centering
    \includegraphics[width=0.49\textwidth]{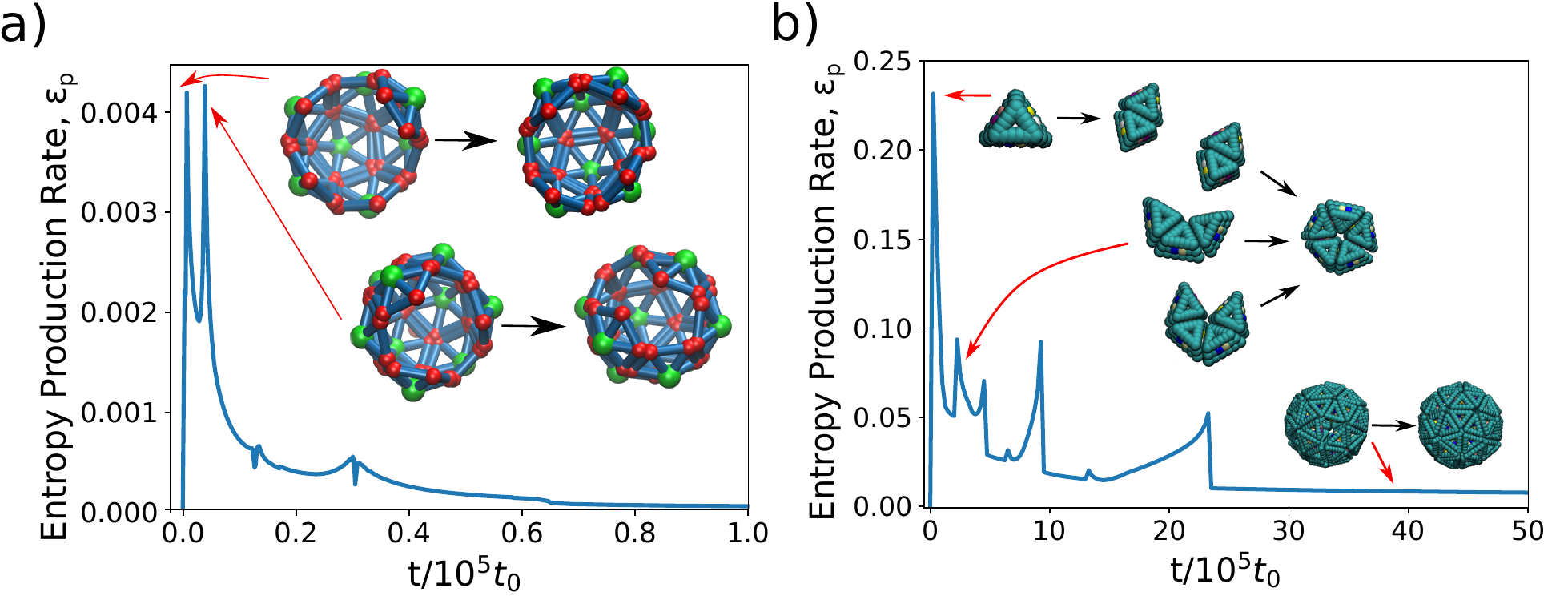}
    \caption{Entropy production rates as a function of time as computed by the MultiMSM for \textbf{(a)} dodecahedron assembly with $\Ebind=5.5$ and $\ctot=0.0156$ and 
    \textbf{(b)} T=3 capsid assembly with $(\Eone, \Ethree) = (11,8)$ and $\ctot=1.7\times 10^{-4}$. Selected peaks and other areas of interest are labeled with the transition or transitions that contribute most to the total entropy at that time. 
    }
    \label{fig::entropy_production}
\end{figure}

Figs.~\ref{fig::entropy_production}a,b show the result of computing the entropy production rate as a function of time using the MultiMSM for dodecahedron assembly at the intermediate binding energy, $\Ebind=5.5$, and the T=3 model. 
For the dodecahedron case, we observe an initial spike in the entropy production rate at early times, corresponding to a non-equilibrium flux of monomers into larger intermediates. The entropy production rate then decreases rapidly and remains small throughout the remainder of the simulation. Importantly, note the difference in timescales for the entropy production rate decreasing toward zero and the yields reaching a steady state in Fig.~\ref{fig::dodec_all_yields}b. This suggests the assembly is near-equilibrium once the monomer concentration reaches roughly half its starting value. This is consistent with the accepted notion that productive assembly occurs when the system is near equilibrium \cite{Zlotnick2007, Hagan2014, Zandi2020}.

We can also track the dominant contributions to the sum in Eq.~\eqref{eq::entropyMultiMSM} to better understand the assembly process. The two initial peaks are labeled with a representation of the transition contributing the most to the entropy at that time. The first peak corresponds to the transition between $(\nsub, \bonds) = (7,12)$ to $(\nsub, \bonds) = (8,15)$, accounting for approximately $7\%$ of the entropy production rate, while the second peak corresponds to the transition between $(\nsub, \bonds) = (9,18)$ to $(\nsub, \bonds) = (10,21)$, accounting for approximately $11\%$ of the entropy production rate. These two transitions are dominant considering that the next largest contributions account for less than $2\%$. 
This observation suggests these are the dominant nearly irreversible transitions in the assembly process, and indeed we can see these transitions occur on the most likely assembly pathway identified in Fig.~\ref{fig::tpt}b. The enhanced lack of reversibility for these transitions can be understood because both transitions add three bonds to the configuration, compared to just one or two for most of the other early transitions along the pathway. This result demonstrates that the entropy production can provide important insights into a reaction pathway by identifying the key transitions that stabilize intermediates or products.
 
The $T=3$ model (Fig.~\ref{fig::entropy_production}b) exhibits a similar large spike in entropy production at early times due to rapid formation of dimers and small intermediates, followed by a decay over time.
The initial peak corresponds to the monomer to dimer transition, accounting for approximately half the entropy production at this time. 
The next peak is dominated by transitions from a dimer, trimer, or tetramer to a pentamer, with each transition contributing approximately $2.5\%$ for a total of $7.5\%$. These are the expected dominant contributions, as the strong intra-pentamer interactions make pentamerization a nearly irreversible process under these parameter values. 
There are no intermediate peaks that have a dominant contribution to the total entropy production, but rather many transitions with roughly the same contribution. 
An interesting observation for this example is that the entropy production rate does not tend to zero at large times, but rather decays to a small positive number, approximately $0.006$. It does not decay further even if we increase the final time by up to two orders of magnitude. This entropy production is almost entirely ($>99\%$) due to configurations with $59$ subunits transitioning to the T=3 capsid, as well as configurations with $60$ subunits but the wrong bond configuration transitioning to the T=3 capsid. At this point in the assembly, the free monomer concentration has already decayed to less than $1\%$ of its initial value, which indicates that the system has become trapped in metastable states. This result demonstrates that entropy production provides a useful measure of the extent to which a system is kinetically trapped.

\subsection{Free Energy}
\label{sec::free_energy}

\begin{figure*}[ht!] 
    \centering
    \includegraphics[width=0.98\textwidth]{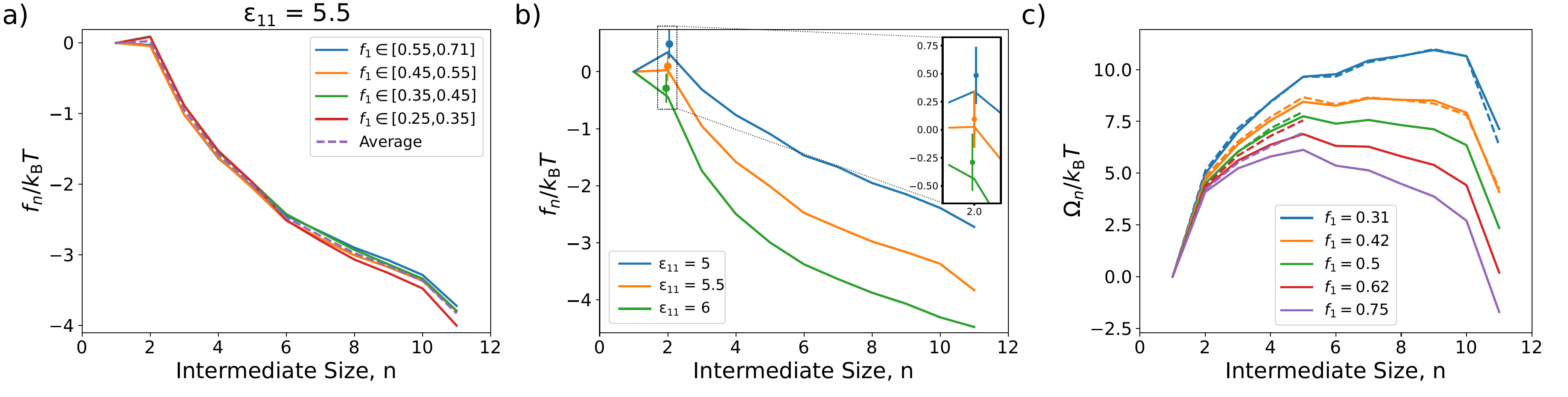}
    \caption{MultiMSM free energy calculations for dodecahedron assembly. 
    \textbf{(a)} The per-subunit Helmholtz free energy $f_n$  computed from each component MSM as a function of intermediate size $n$ for $\Ebind=5.5$ (solid lines), showing that the results are independent of monomer concentration. The free energy averaged over the component MSMs is shown as a dashed line. 
    \textbf{(b)} The averaged Helmholtz free energy profiles as a function of binding energy, $\Ebind$. The inset compares the free energy for the dimer computed from the MultiMSM against the estimated value and standard error obtained directly from dynamical simulations (without using an MSM) (see supplement Section \ref{SMsec::dimerization_fe}). 
    \textbf{(c)} The grand free energy, computed analytically from the Helmholtz free energy using Eq.~\eqref{eq::grand_free_energy} and evaluated at the average monomer fraction within each discretization bin (solid lines). Dashed lines show the grand free energy computed directly from the MSM equilibrium distribution, for monomer fractions and intermediate sizes that are sufficiently close to equilibrium to enable comparison. 
    }
    \label{fig::free_energy_dodec}
\end{figure*}

Another useful application of MSMs is calculating the equilibrium free energy of each microstate, which can then be projected onto reaction coordinate(s) for mechanistic insight. Notably, the MultiMSM allows computing the \textit{equilibrium} free energy profile independent of free monomer concentration, from nonequilibrium simulations at arbitrary monomer concentration.

While a typical MSM enables computing the Helmholtz free energy from the equilibrium distribution of microstates $\vec{\pi}$ as $F_i=-\kt\log(\pi_i)$, in the MultiMSM framework the transition matrix depends on the free subunit concentration and thus the corresponding chemical potential. Therefore, the MultiMSM gives the grand free energy for the $j$-th component MSM as \begin{equation}
\label{eq::grand_free_energy}
    \Omega_i^j = -\kt\log(\pi_i^j) = F_i - \mu_j (n_i-1),
\end{equation}
where $\Omega_i^j$ is the grand free energy for state $i$ in the $j$-th MSM, $\mu_j = \kt\log(c_j/\css)$ is the chemical potential for the $j$-th MSM with $c_j$ as the average monomer concentration within the bin and $\css$ as the standard state concentration, and $n_i$ is the number of subunits in microstate $i$. 

Importantly, since the Helmholtz free energy $F_i$ in Eq.~\eqref{eq::grand_free_energy} is independent of the chemical potential (and thus the free monomer concentration), it should be the same for each component MSM of the MultiMSM. Therefore, its statistics can be improved by averaging over each of the component MSMs. 

To simplify the following presentation, we project the free energy onto a reaction coordinate where there is one state per intermediate size $n$, but the approach readily generalizes to multiple states per size.
To compute $F_i$ within a particular MSM, we compute the equilibrium distribution from the transition matrix and then the set of equilibrium constants for the formation of each $n$-mer, 
\begin{equation}
\label{eq::equilibrium_constants}
    K_n = \left[ n\right]/\left[ 1 \right]^n,
\end{equation}
where $\left[n\right]$ denotes the equilibrium concentration of $n$-mers. Note that any absorbing states (states or groups of states for which there were an insufficient number of exit transitions to estimate an outward transition rate) should be eliminated from the transition matrix before computing equilibrium quantities. As noted above, outward transition rates could be computed for such states by combining free energy calculations with the dynamical simulations used to estimate transition rates \cite{Trendelkamp-Schoer2016}, but we have not implemented this approach for the present work. Then, the concentration of subunits in absorbing states should be subtracted from the total subunit concentration to ensure proper normalization of the grand free energy (see supplement Sections \ref{SMsec::absorbing_states} and \ref{SMsec::equilibrium_constants} for details). 

The Helmholtz free energy can be computed as
\begin{equation}
    \label{eq:helmholtz_free_energy}
    F_n = -\kt\log(\css^{n-1} K_n).
\end{equation}
While the equilibrium constant corresponds to the true observable and is thus independent of standard state, the free energy values necessarily depend on $\css$. For the results presented here, we choose $\css$ in simulation units such that there is one subunit per circumscribed volume occupied by the subunit: $\css=0.66/\lzero^3$ for the pentagonal subunit and $\css=0.17/\lzero^3$ for the triangular subunits.

\begin{figure}[ht!] 
    \centering
    \includegraphics[width=0.49\textwidth]{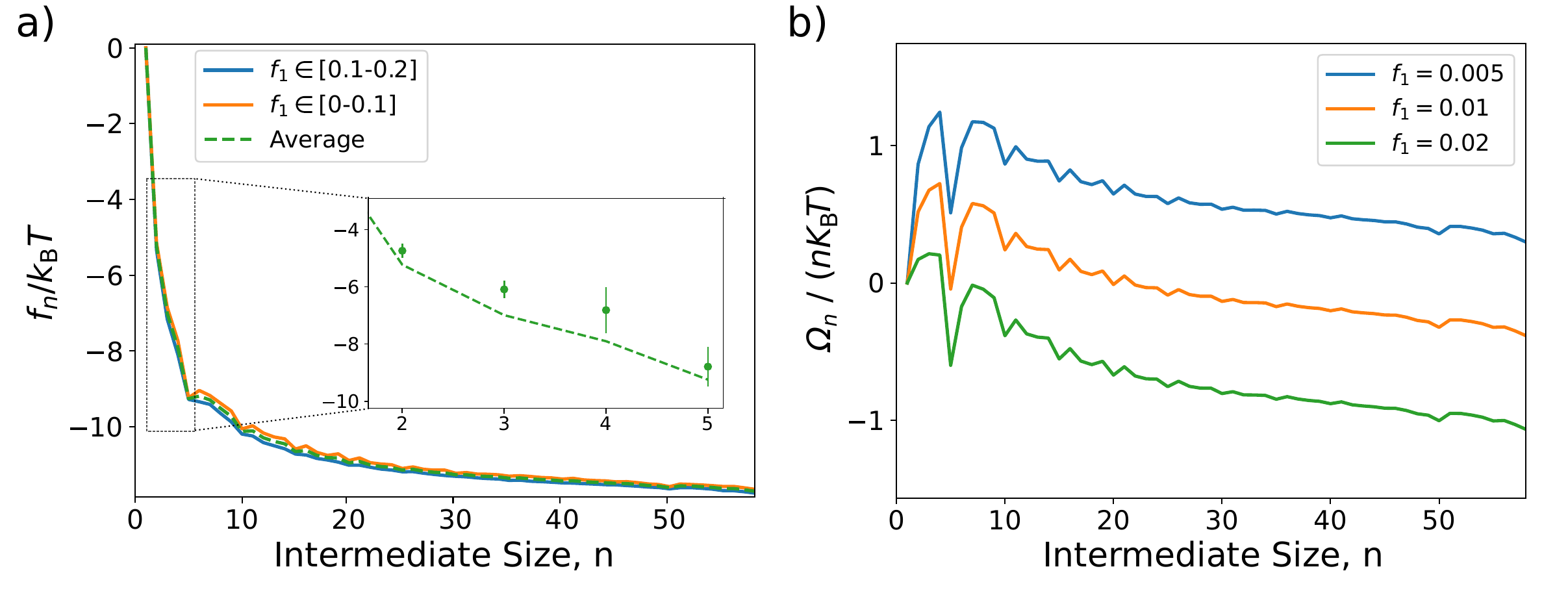}
    \caption{MultiMSM free energy calculations for $T=3$ capsids. 
   \textbf{(a)} The per-subunit Helmholtz free energy for each component MSM (solid lines) and the averaged profile (dashed) line, showing that the results are independent of monomer concentration.      The inset compares the free energy estimates  from the MultiMSM against results from dynamical simulations up to pentamers (see supplement Section \ref{SMsec::pentamerization_fe}).
    \textbf{(b)} The per-subunit grand free energy, computed analytically from the Helmholtz free energy using Eq.~\eqref{eq::grand_free_energy} and evaluated for several free monomer fractions. 
    }
    \label{fig::free_energy_triangles}
\end{figure}

Figs.~\ref{fig::free_energy_dodec}a and \ref{fig::free_energy_triangles}a show the Helmholtz free energy per subunit $f_n=F_n/n$ as a function of intermediate size, computed from the MultiMSM using Eq.~\eqref{eq:helmholtz_free_energy}, across the monomer fraction bins for dodecahedron assembly with $\Ebind=5.5$ and $T=3$ capsid assembly, respectively. We see that the curves from different component MSMs collapse, showing that the Helmholtz free energy is independent of monomer concentration as it should be. We then average this data to construct the best statistical estimate. As expected, the barrier height decreases and the stability of each intermediate increases with increasing binding affinity. As an independent test, we compare the dimer Helmholtz free energy against the free energy computed directly from simulations (without using MSMs) in the inset (see supplement Sections \ref{SMsec::dimerization_fe} and \ref{SMsec::pentamerization_fe} for details).  For the $T=3$ system we performed the same procedure and compared against direct simulations for all intermediates up to a pentamer, which also show strong agreement in Fig.~\ref{fig::free_energy_triangles}a. 

The Helmholtz free energy enables computing the grand free energy at any free monomer concentration by adding the chemical potential term as in Eq.~\eqref{eq::grand_free_energy}. 
For example, Figs.~\ref{fig::free_energy_dodec}c and \ref{fig::free_energy_triangles}b show the grand free energy for dodecahedra with $\Ebind=5.5$ and $T=3$ capsids, respectively, at several representative subunit concentrations. 
For the dodecahedron system, we evaluate the grand free energy at monomer fractions corresponding to the average monomer fraction within each discretization window (solid lines). This allows for a direct comparison of our computed grand free energy with the result of computing it directly from the component MSM equilibrium distribution (dashed lines). Note that the nucleation barrier increases, and the stability of intermediates decrease, as free monomer concentrations decrease due to the increased monomer chemical potential. 
Note that these curves should coincide only when the assembly is at or near equilibrium. Fig.~\ref{fig::entropy_production}a shows that the entropy production vanishes once the monomer fraction is below roughly $0.5$, and we see excellent agreement between the two calculations at lower monomer fractions. 
For larger monomer fractions, we may expect structures smaller than the critical nucleus to be in quasi-equilibrium \cite{Hagan2014}, so we make comparisons up to an intermediate size of $5$. In this case, we find loose agreement between the curves that improves as the monomer fraction decreases. 

For $T=3$ capsids, we do not make comparisons with the equilibrium MSM as we have previously determined that this example is kinetically trapped and thus not in equilibrium. We instead evaluate this for the range of monomer fractions that we observe at the final simulation time of our long trajectories. Interestingly, these profiles exhibit local minima at multiples of $5$, reflective of the pentamer biased assembly parameters and consistent with previous computational and experimental results \cite{Wei2024}.

\section{Conclusions}
\label{sec::conclusions}
We have described the MultiMSM, a general framework to construct MSMs for systems in which many clusters assemble simultaneously and concentrations (and potentially other parameters) change over the course of the reaction. Using two model systems, we show that MultiMSMs can accurately describe assembly dynamics over the long timescales required to approach equilibrium, even when constructed from trajectories that are orders of magnitude shorter. This capability enables particle-based simulations with complex models to be simulated at experimentally relevant concentrations. The degree of speed up enabled by the MultiMSM increases exponentially for systems with large nucleation barriers, and the method is well-suited for systems that assemble by diverse pathways. Moreover, the method is trivially parallelizable and thus highly scalable.

In addition to extending on previous work by allowing for multiple clusters and the depletion of free subunits, the MultiMSM approach allows for a number of further applications. 
Notably, a MultiMSM constructed at one total subunit concentration can be used to perform an approximate parameter sweep over a wide range of concentrations with no additional sampling. This capability corresponds to orders of magnitude additional speed up in comparison to brute-force simulations. The results are highly accurate for conditions leading to productive assembly, under which concentrations of intermediates remain relatively low, and qualitatively accurate for more aggressive assembly conditions that lead to a buildup of intermediates. 
The MultiMSM framework computes transition path theory quantities, such as the committor probability describing the extent of progress along a reaction coordinate and the relative flux along different assembly pathways. 
It can also be used to estimate the Helmholtz free energy of a system, as well as compute the grand free energy as a function of the monomer concentration. 
Further, the method allows for efficient calculation of entropy production rates, a quantity which has been difficult to compute from particle-based simulation trajectories. We find that the entropy production rate provides a useful quantification of how far an assembly reaction is from equilibrium and whether it is susceptible to kinetic traps. More interestingly, by analyzing which transitions contribute most to the entropy production rate, one can identify factors that stabilize critical nuclei or other key intermediates, and whether these transitions lead to productive assembly or engender kinetic traps. 
These insights can form the basis for rational design of synthetic assembly systems, or, in the case of biomedically relevant assembly systems such as viruses, help to identify targets for antiviral molecules that interfere with assembly \cite{Kim2021,Kondylis2018,Schlicksup2018, Pavlova2022, Wang2023, Kra2023}.

\begin{acknowledgments}
This work was supported by the NSF through DMR 2309635 and the Brandeis Center for Bioinspired Soft Materials, an NSF MRSEC (DMR-2011846). We also acknowledge computational support from NSF XSEDE computing resources allocation TG-MCB090163 (NCSA Delta GPU) and the Brandeis HPCC which is partially supported by the NSF through DMR-MRSEC 2011846 and OAC-1920147.
\end{acknowledgments}

\section*{Conflicts of Interest}
The authors declare no conflicts of interest. 

\section*{Data Availability Statement}
Our Python library to post-process the results of a HOOMD simulation \cite{Trubiano_SAASH_2024} and our Python library to construct and perform calculations on MultiMSMs \cite{Trubiano_MultiMSM_2024} is freely available on Github. 
Our simulation scripts, a subset of post-processed trajectory data, analysis scripts, and figure generating scripts are hosted on the Open Science Framework OSFHome (https://osf.io/ak49n/).

\bibliographystyle{apsrev4-1}
\bibliography{references}

\clearpage


\setcounter{figure}{0}
\setcounter{section}{0}

\makeatletter
\NewDocumentCommand{\MakeTitleInner}{ +m +m +m }{
    \newpage%
    \null%
    \vskip 2em%
    \begin{center}%
        \let \footnote \thanks
        {\LARGE #1 \par}
        \vskip 1.5em%
        {%
            \large
            \lineskip .5em%
            \begin{tabular}[t]{c}%
                #2
            \end{tabular}\par%
        }%
        \vskip 1em%
        {\large #3}
    \end{center}%
    \par
    \vskip 1.5em%
}
\NewDocumentCommand{\MakeTitle}{ +m +m +m }{%
    \begingroup
        \renewcommand\thefootnote{\@fnsymbol\c@footnote}%
        \def\@makefnmark{\rlap{\@textsuperscript{\normalfont\@thefnmark}}}%
        \long\def\@makefntext##1{\parindent 1em\noindent
            \hb@xt@1.8em{%
                \hss\@textsuperscript{\normalfont\@thefnmark}%
            }##1%
        }%
        \if@twocolumn
            \ifnum \col@number=\@ne
                \MakeTitleInner{#1}{#2}{#3}
            \else
                \twocolumn[\MakeTitleInner{#1}{#2}{#3}]%
            \fi
        \else
            \newpage
            \global\@topnum\z@   
            \MakeTitleInner{#1}{#2}{#3}
        \fi
        \thispagestyle{plain}\@thanks
    \endgroup
    \setcounter{footnote}{0}%
}
\makeatother

\makeatletter
\renewcommand \thesection{S\@arabic\c@section}
\renewcommand\thetable{S\@arabic\c@table}
\renewcommand \thefigure{S\@arabic\c@figure}
\renewcommand \theequation{S\@arabic\c@equation}
\makeatother

\counterwithin{figure}{section}

\onecolumngrid
\MakeTitle{Markov State Model Approach to Study Self-Assembly - Supplement}
          {Anthony Trubiano and Michael Hagan}
          {Martin Fisher School of Physics, Brandeis University, Waltham, Massachusetts 02454, USA}

\section{Simulation Details}
\label{SMsec::sim_details}
\subsection{Dodecahedron system}

The subunits are rigid bodies composed of several kinds of pseudoatoms. There are attractor pseudoatoms (`A') at the vertices of a regular pentagon, which facilitate subunit assembly via an attractive Morse potential with an equilibrium length of $\Lzero = 0.2$, a range parameter $\alpha = 2.5/\Lzero$, a cutoff distance of $2$, and a well-depth (subunit-subunit binding strength) $\Ebind$ that can be varied. In this work, we use values $\Ebind \in \{5.0, 5.5, 6.0\}$. 
We also include a top pseudoatom (`T') and a bottom pseudoatom (`B'), located at positions $z=\pm 0.5$ with respect to the center of the regular pentagon in the xy plane. The `T' pseudoatoms interact with other `T' pseudoatoms via a repulsive Lennard-Jones potential with $\Stt=2.1$, cutoff distance equal to $\Stt$, and well-depth $\Ett = \Ebind/4$. These values favor a subunit-subunit binding angle consistent with that of a dodecahedron. 
The `B' pseudoatoms have a similar repulsive interaction with `T' pseudoatoms, with $\Stb=1.8$, a cutoff distance equal to $\Stb$, and well-depth $\Etb = \Ebind/4$. This interaction helps to prevent upside down assembly, i.e. ensuring that the top atom is in the direction of the outward normal vector. 
Finally, we add edge (`E') pseudoatoms at the midpoint between each adjacent vertex. These have no interactions and thus do not affect the simulation, but they are used to track assembly progress more easily compared to vertex pseudoatoms. 

Results are reported in units for which the unit length $\lzero$ corresponds to the edge length of the pentagonal subunit and energies are measured in units of $\kt$. 
Simulations are initialized with $125$ pentagonal subunits, enough to form $10$ dodecahedral capsids ($12$ subunits each), on an equally spaced lattice. Subunit positions and orientations are then equilibrated with a purely repulsive potential for $2\times 10^5$ time steps before writing any output. 
The simulation box is a cube with periodic boundary conditions and side lengths $20\lzero$, giving a total subunit concentration of $\ctot=0.0156/\lzero^3$. The time step is $0.001\tzero$, and the base simulations to construct MSMs are run for $5\times 10^6$ time steps unless otherwise specified. The simulations use the HOOMD-blue \cite{Anderson2020} version 3.9.0 Langevin integrator, with an inverse temperature $\beta = 1$. 
The configurations are logged every $\Delta t = \tzero$ units of simulation time.

\subsection{$T=3$ capsid system}
 The subunits are rigid triangles with each edge consisting of three stacked layers of six overlapping `excluder' pseudoatoms at a specified bevel angle. These excluder pseudoatoms interact with all other pseudoatoms in the simulation through a Weeks-Chandler-Anderson (WCA) potential to enforce excluded volume. Embedded in the middle row of each edge of the triangle are two attractor pseudoatoms, which have attractive Lennard-Jones interactions with the pseudoatoms on complementary edges to facilitate edge-edge binding of the subunits. Each pseudoatom has the same diameter, $\sigma$, which we set as the unit distance $\lzero$. To match the dimensions of the experimental subunit \cite{Wei2024}, we can set $\sigma=18$nm in real units. 

As described so far, the triangular subunits can be designed to form a broad variety of target structures by tuning the interaction strengths, side lengths, and bevel angles. For a $T=3$ capsid target, we make two sides of the triangle equivalent. Sides 1 and 2 have an edge length of $3\sigma=54$nm and have complementary pseudoatoms (Side 1 has pseudoatoms `4' and `5', which bind with pseudoatoms `7' and `6' on Side 2, respectively) that attract with a binding energy of $\Eone$. These two sides do not interact at all with Side 3, which has a slightly longer edge length of $3.35\sigma=60.3$nm, with attractive pseudoatoms (pseudoatoms `2' and `3') that are self-complementary with a binding well-depth of $\Ethree$. The bevel angle of each side is the same, approximately $11.64^\circ$. This design can produce hierarchical assembly \cite{Wei2024}. When $\Ethree\gg\Eone$, dimers form rapidly via the Side 3 -- Side 3 interaction, and then the dimers more slowly assemble into larger structures via the Side 1 -- Side 2 interaction. Conversely, when $\Ethree\ll\Eone$, pentamers form rapidly and then subsequently assemble into larger structures. In this work, we consider an example of pentamer-biased assembly, with $\Eone=11$ and $\Ethree=8$. 

The simulations contain $N=600$ triangular subunits, enough to form $10$ $T=3$ capsids ($60$ subunits each). 
The simulation domain is an $L\times L\times L$ box with periodic boundary conditions, whose side lengths determine the total subunit concentration, $C = N/(\NA L^3)$, where $\NA$ is Avogadro's number. 
We set $C=50$nM to be on the order of experimental conditions, for which the corresponding box side length is $L=2.71$ microns. 
In simulation units, the box side lengths are $153\lzero$, giving a total subunit concentration of $\ctot=1.7\times 10^{-4}/\lzero^3$. 

We initialize subunit positions on an equally spaced, truncated lattice. Subunit positions and orientation are then equilibrated with a purely repulsive potential for $8\times 10^4$ time steps, after which attractive interactions are turned on. We use a time step of $0.0025\tzero$, and the base simulations to construct MSMs are run for $5 \times 10^8$ time steps unless stated otherwise.


\subsection{Bond Definitions and Discrete States}
\label{SMsec::bond_definition}

For the dodecahedron system, the pentagonal subunits have edge (`E') pseudoatoms that align when the adjacent vertex attractors bind with another subunit. We use a cutoff distance of $0.3\lzero$ between these edge pseudoatoms to define a bond. 
For the $T=3$ capsid assembly, there are a pair of complementary pseudoatoms on Sides 1 and 2 (`4' and `7' pseudoatoms), and Side 3 is self-complementary (`2' and `3' pseudoatoms). We use a cutoff distance of $1.3\lzero$ for both pairs of complementary pseudoatoms in this case. 
These values were chosen by selecting the minimum distance that would correctly identify fully formed capsids across many different realizations of the assembly. Our results are insensitive to increasing these cutoffs up to the next-nearest neighbor bond distances, which are approximately $0.8\lzero $ for the pentagonal subunit and $2.4\lzero $ for the triangular subunits. 
Decreasing these cutoffs results in frequent oscillations in the number of bonds, even for stable configurations like the fully formed dodecahedron. Such behavior is undesirable when building a Markov state model, as it overestimates the rate of bond breakage in the model, so we ensure our choice of cutoff correctly identifies the target structure over long timescales at the strongest binding energies. 

Using this definition of a bond, we construct our discrete state space as all observed combinations of $(\nsub, \bonds)$, where $\nsub$ is the number of subunits in a cluster and $\bonds$ is the distribution of the number of bonds in a cluster. Defined in this way, the number of unique discrete states we observe is system dependent. 
There are $180$ states for dodecahedral capsid assembly with binding energies $\Ebind=5$ and $\Ebind=5.5$. 
For $\Ebind=6$, the interactions are strong enough to result in transient intermediates that are larger than $12$ subunits, resulting in a state space size of $447$. Despite the enhanced size, most of these states are extremely low probability and could likely be excluded without affecting the model predictions, though we have not tested this. 
For $T=3$ capsid assembly, the number of states is much larger due to a larger capsid size and having two distinguishable bond types. For the pentamer-biased assembly set that we consider, the state space size is $1662$. Since the hierarchical assembly pathways likely decreases the number of intermediates, the state space might be larger for other parameter sets.

\section{Transition Counting Example}
\label{SMsec::transition_counting}

\begin{figure}[ht!] 
    \includegraphics[width=0.80\textwidth]{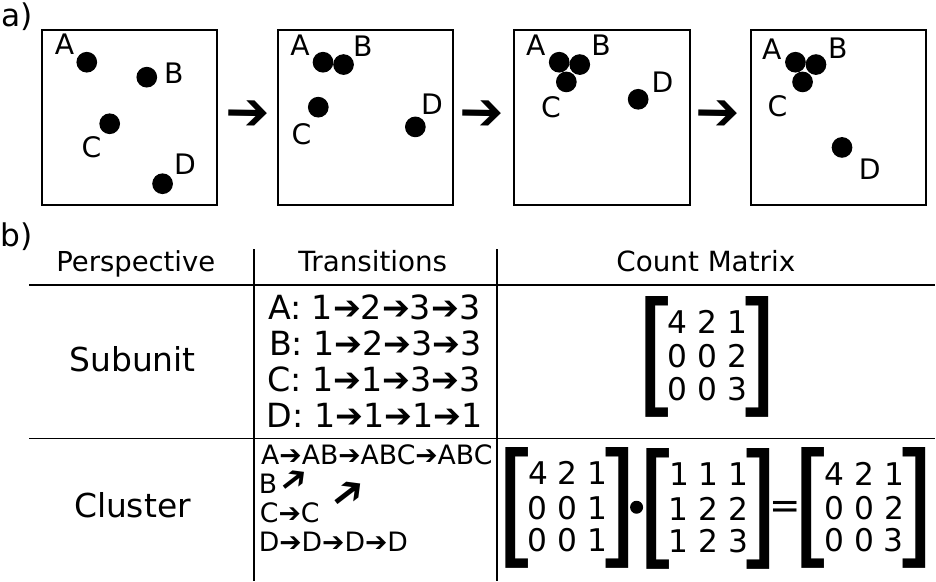}
    \caption{A simple assembly example showing two equivalent approaches to count transitions. 
    \textbf{(a)} A simple example time series of four disk shaped particles with attractive interactions undergoing clusterization. 
    \textbf{(b)} A table detailing how to count transitions from both a subunit and cluster perspective for this example. In the subunit perspective, a time series of the cluster size containing each subunit is computed, and a count matrix is built from the number of times each transition is observed. In the cluster perspective, only merging and splitting of clusters is tracked. The count matrix is built by tallying the number of times a cluster of size $i$ points to a cluster of size $j$, and then multiplied by $W_{ij} = \min(i,j)$ to get the counts in the subunit perspective.}
    \label{SMfig::counting}
\end{figure}

As stated in the main text, the construction of the MSM count matrix from the set of cluster transitions must maintain mass weighting. There are two equivalent ways to count transitions such that the mass weighting is preserved. 
The first is to take a subunit perspective; for each individual subunit, compute the time series of its cluster size and count every transition that occurs. Since this procedure loops over each subunit in a given cluster, it automatically accounts for the mass-weighting. This is at the cost of efficiency however, since each subunit in a cluster stores the same transition information. 
Instead, we take a cluster perspective, which is described in Section \ref{sec::TransitionCounting}. Here, each cluster transition is only counted a single time and then weighted in the count matrix by the number of subunits in the smaller cluster. Since each entry of the count matrix will be weighted by the same number, we can actually count all of the unweighted transitions first, then apply the weight matrix $W_{ij} = \min(\nsub_i,\nsub_j)$, where $i$ and $j$ are generic state indices.  
This approach is more efficient both in terms of construction time and memory usage. 

Fig.~\ref{SMfig::counting} shows a simple example of this counting procedure, using a lag of one frame. Every transition is enumerated for each perspective, and we see that the same count matrix is obtained from both approaches.  In this example, states are defined as the number of subunits in a cluster.

\section{A Note on Solving the Backward Equation}
\label{SMsec::backward_eq}
While not directly relevant for the main text of this paper, for some applications it may be useful to solve the backward Kolmogorov equation for the MultiMSM as well, and there are some subtleties to be aware of. For reference, the backward equation for a standard MSM can take the form
\begin{equation}\label{SMeq::BKE}
    \niceVec{f}{n} = \mat{P} \niceVec{f}{n+1}
\end{equation}
where the column vector $\niceVec{f}{n}$ can be interpreted as the conditional expectation of some reward function of the final state of the Markov chain at time $n$. This equation is prescribed as a final condition, $\niceVec{f}{N}$, and solved backwards in time. This is straightforward for the standard MSM because the transition matrix is fixed for all time. This is not the case for the MultiMSM; we need to store the sequence of transition matrices used to solve the forward equation to know what transition matrix to use at each step of solving the backward equation. Formally, we need to solve
\begin{equation}\label{SMeq::BKE_multi}
    \niceVec{f}{n} = \mat{P}_{m^n} \niceVec{f}{n+1}, \quad m^n = \text{index}(p_0^n).
\end{equation}
This raises two subtleties about solving the backward equation for the MultiMSM. First is that the forward equation must be solved first to know the $p_0^n$ for all times. Second is that, if the smoothing operation described in the main text is used while solving the forward equation, the exact linear combination of transition matrices used at each time step must also be saved. 

To efficiently address these issues, we keep track of how the transition matrices are constructed during the forward solve. If no smoothing is used, we simply store the index of the transition matrix used at each time step to be referenced later. If smoothing is used, we store a list of tuples, where the first entry stores the index of the matrix and the second entry stores the weight. If in a transition region, there will be two indices with their respective weights that can be used to reconstruct that transition matrix. If not in a transition region, there will be only one index with a weight of one.

\section{Choosing a Good Monomer Fraction Discretization}
\label{SMsec::discretization_opt}

In this section we describe a protocol that aims to select an optimal monomer fraction discretization, and a metric to gauge the quality of the discretization.   

Our cluster analysis software, in addition to returning all transitions a particular cluster makes along its lifetime, can also be used to track the yield of any of the discrete states, $s$, from any set of simulations. For each system, we perform a set of simulations that are initialized with all subunits as monomers with thermalized positions and orientations. We average the mass-weighted yields of each discrete state over each of these simulations to get an estimate of the true probability of observing that structure, $\{\hat{p}_s^n\}_n$. In particular, we can do this for `important' intermediate states, such as monomers, dimers, pentamers, the full capsid, other intermediates that are relatively high probability or are involved in important transitions, or any combination thereof. We can then compare these curves to the corresponding entries of the time-dependent probability distribution we get from solving Eq.~\eqref{eq::FKE_multi} from the MultiMSM with a discretization $D$, $\{p_s^n(D)\}_n$. Our metric for the quality of the discretization is then a normed difference of these two quantities,
\begin{equation}
    C(D) = \| \hat{p}_s - p_s(D) \|^p = \left( \sum_n \abs{\hat{p}_s^n - p_s^n(D)}^p \right)^{1/p},
\end{equation}
where we typically use the $2$-norm, but leave this as a tunable parameter for the optimization. 

This metric can then be used as the cost function to minimize for some optimization procedure in which the discretization points $d_i$ are varied. We experimented with both Monte Carlo and gradient descent optimization schemes, but both tended to be both slow and prone to getting stuck in local minima. Instead, we apply a sequential optimization procedure that works as follows. 

Consider the discretization $D_N = (0, d_1, d_2, \cdots, d_N, 1)$ as an initial guess. Since all our examples begin with a monomer fraction of $1$ and deplete monomers over time, the bins closer to $1$ have a larger effect on the accuracy of the MultiMSM yield curves. This is because any error made in the bin closest to $1$ will propagate to all future bins as the monomers deplete. Therefore, our sequential optimizer works by fixing $d_1$ through $d_{N-1}$ and choosing the optimal value for $d_N$, that minimizes the cost function. We then vary $d_{N-1}$ while keeping all of the other bins fixed, then $d_{N-2}$, and so on, continuing until we reach $d_1$. It is possible that modifying the other bin locations has shifted the optimal value for $d_N$, so we perform another sequential optimization cycle. We repeat the procedure until we reach a cutoff number of cycles, or until a cycle terminates without changing any of the $d_i$. 

Every time a new discretization is tested the MultiMSM must be reconstructed since many of the transitions may now lie in a different bin. While we have implemented a caching system to reduce the time it takes to reconstruct the MultiMSM with a new discretization, the construction time and time to solve the forward equation is non-negligible. Therefore, we want to minimize the number of discretizations we test while performing this optimization. When optimizing $d_i$, we construct the interval between neighboring discretization points, $[d_{i-1},d_{i+1}]$, and place $M$ equally spaced points within this interval. We keep $M$ relatively small, typically $M=4$, and check if using those value for $d_i$ reduces the cost function. If not, we keep the same $d_i$ and move to the next point. This keeps the optimization time to a reasonable level; typically we observe convergence in about $~5{-}10$ minutes, using $N=6{-}8$, $M=4$, and $5$ cycles on a single $3.5$ GHz CPU.

\section{Additional Adaptive Sampling}
\label{SMsec::adaptive_sampling}

A powerful feature of MSMs is the ability to simulate dynamics on timescales that are much longer than those of the brute force dynamics simulations that are used to estimate the transition matrix elements. Adaptive sampling, in which sampling is focused on important transitions, increase the accuracy of such predictions.

Ensuring the accuracy of longtime predictions is more challenging for the MultiMSM than a standard MSM because the transition matrices change over time in the MultiMSM. Straightforward brute-force sampling may not result in good statistics for the long time behaviors. For example, in the pentamer-biased simulations of the $T=3$ capsid, the average monomer fraction at the final simulation time is about $5\%$. 
About $1/4$ of the base trajectories sampled lower values than this, but there are only $\sim 20000$ transitions sampled for $0\% < \mfrac < 3.5\%$ while there are $\sim 160000$ transitions sampled for the smaller interval $3.5\% < \mfrac < 6\%$. Thus, MultiMSM dynamics on timescales that lead to such low monomer fractions will have limited accuracy without additional sampling at low monomer fractions.
We have developed techniques to more efficiently generate such data, particularly in typical challenging cases.

The first case, mentioned above, occurs when the monomer fraction at the final simulation time has not yet reached its equilibrium value, but is close to it. To generate additional sampling at lower monomer fractions, we identify which of the existing trajectories ended with the lowest monomer fraction and initialize a new simulation in the final frame of the existing one. Since transition rates will depend on the precise distribution of intermediates present, we try to perform this for as many different starting frames as possible, with multiple random seeds for each. We refer to this as `continued' sampling. 
A particularly challenging sub-case occurs when the interaction strengths are relatively weak. In this case, the nucleation of a cluster is a rare event, and reaching equilibrium may take orders of magnitude longer than available computational times. In this case, we artificially push the system closer to equilibrium, in a way that does not bias the equilibrium distribution of intermediates. Since the equilibrium distribution in assembly systems with a large nucleation barrier can be well approximated by a coexistence of just full capsids and monomers \cite{Zlotnick1994,Hagan2014,Zandi2020}, we can construct a starting frame from an existing frame by randomly selecting monomers and manually assembling them into a capsid, placing them in the simulation box in such a way that there is no overlap. We refer to this as `fraction' sampling. Alternatively, one can also remove the monomers forming full capsids from the simulation box. As long as the simulation box size is reduced to account for the volume of the capsid, this should not bias the dynamics of the remaining monomers, and could significantly speed up the simulations depending on the system size. We refer to this as `reduced' sampling. 

Another issue can arise in cases where the interaction strengths are relatively strong. In this case, monomers will very quickly deplete, resulting in very few sample transitions being used to construct MSMs in the discretization bins corresponding to larger monomer fractions. In this case, we perform many short simulations, anywhere from $5$ to $10$ percent of the original simulation time, to gather more samples for the larger monomer fraction values. We refer to this as `short' sampling. Fortunately, we have found that in cases of fast depletion, we can choose the discretization such that the entire fast depletion region is contained in a single discretization bin, which reduces the need for this additional sampling.

\section{MultiMSM Parameter Overview}
\label{SMsec::parameters}

Here we summarize the parameters used to construct our MultiMSMs for each example in the main text. Relevant parameters include the amount of sampling data and what type of simulation it came from, the monomer fraction discretization, the minimum number of observations of a transition, and the smoothing parameter used to solve the forward equation. 

First, we list the monomer fraction discretization for each example. These were 
\begin{align*}
    \text{Dodecahedron, } \Ebind = 6.0&\text{: } [0, 0.05, 0.12, 0.13, 0.23, 0.27, 0.52, 0.59, 0.76, 1.0]\\
    \text{Dodecahedron, } \Ebind = 5.5&\text{: } [0, 0.15, 0.25, 0.35, 0.45, 0.55, 0.71, 0.81, 1.0]\\
    \text{Dodecahedron, } \Ebind = 5.0&\text{: } [0, 0.08, 0.16, 0.31, 0.45, 0.61, 0.75, 0.89, 0.98, 1.0]\\
    \text{T=3 Capsid} &\text{: } [0, 0.035, 0.06, 0.1, 0.2, 0.57, 0.63, 1.0]
\end{align*}
In general, these were the result of applying the optimization scheme described above, using the training data to generate sample estimates to optimize against over computationally accessible timescales. 
In the cases where no simulation data is available, such as taking the dodecahedron estimates out to equilibrium for the smaller binding energies, the bins near one were optimized, and subsequent bins chosen manually such that the total number of transition counts in each were roughly balanced. We find the resulting solutions in these cases to be insensitive to small perturbations in the bin locations ($\pm 0.01$). 

The two scalar parameters to the MultiMSM are the smoothing parameter to the forward solver, $\chi$, and the minimum number of observations to keep a particular transition, which we call the prune tolerance. The latter removes transitions that are rare and unlikely to be important to the dynamics. The default values for these parameters are $\chi=0.25$ and a prune tolerance of $1$, which means no pruning. The only exception for the smoothing parameter is the dodecahedron assembly with $\Ebind=6$, which used $\chi=0.5$. The only exception for the prune tolerance is the dodecahedron assembly with $\Ebind=5.5$, which used a prune tolerance of $2$. 

Finally, we report the amount of sampling performed to build each MultiMSM. The base simulations are described above in Section \ref{SMsec::sim_details}, and simulation lengths are reported here in terms of the base simulation length. For `fraction' and `reduced' sampling, a variable number of these simulations were performed per $0.1$ monomer fraction bin, so we report the number of trajectories per bin. Note that bin $i$ in this case refers to simulations initialized with monomer fraction $0.1(i+1)$. 

For the dodecahedron assembly with $\Ebind=6$, we performed $120$ base simulations and $35$ `continued' simulations that were twice the length of the base simulations. For $\Ebind = 5.5$, we performed $80$ base simulations. We also performed `fraction' and `reduced' simulations that were the same length as the base trajectories. The number of `fraction' simulations was $200$ in bin $3$, $300$ in bin $4$, $100$ in bin $5$, and $40$ in bin $6$. The number of `reduced' simulations was $200$ in bin $2$, $200$ in bin $3$, and $50$ in bins $4$, $5$, and $6$. For $\Ebind=5$, we performed $100$ base simulations. In addition to this, we performed $200$ `reduced' simulations in each bin $1$ through $9$.  
For the $T=3$ capsid assembly, we performed $35$ base simulations, as well as $10$ `continued' simulations that were half the length of the base simulations.

\section{Bootstrapping Procedure}
\label{SMsec::bootstrapping}

\begin{figure}[ht!] 
\centering
    \includegraphics[width=0.99\textwidth]{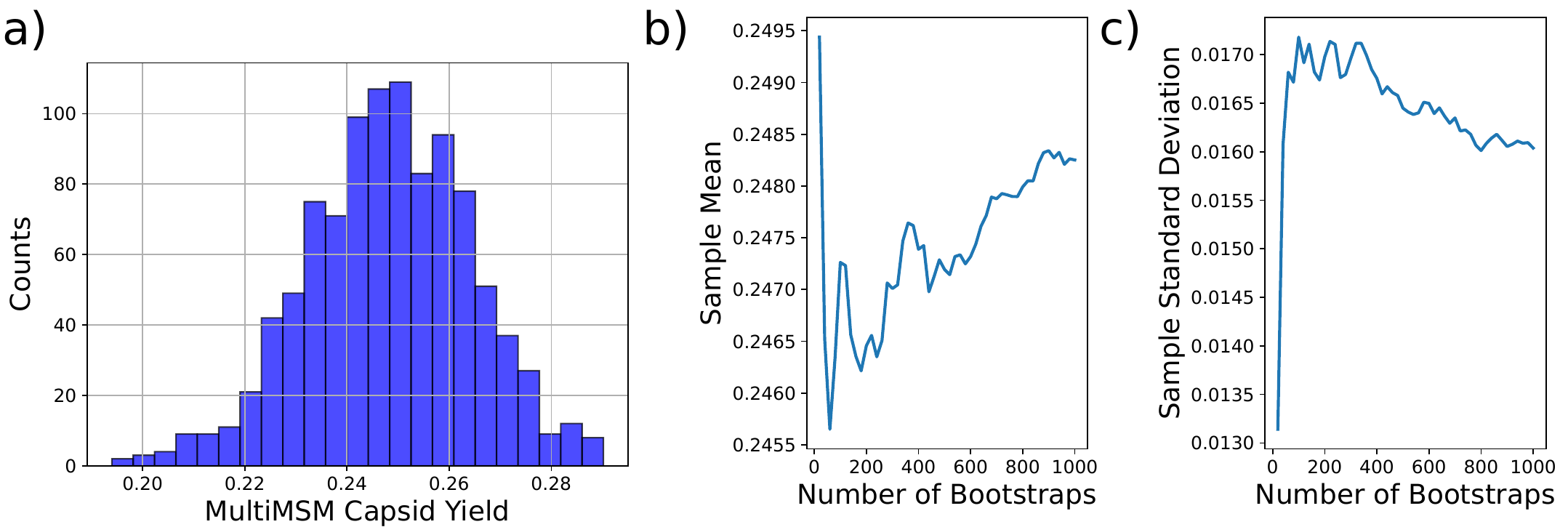}
    \caption{Error estimates by bootstrapping for the MultiMSM capsid yield estimate at the final simulation time for the $T=3$ system. 
    \textbf{(a)} Histogram of MultiMSM capsid yield estimates using $1000$ bootstraps. 
    \textbf{(b)} and \textbf{(c)} show the sample mean and standard deviation, respectively, as a function of the number of bootstraps. 
    }
    \label{SMfig::bootstrapping}
\end{figure}

To estimate errors for the MultiMSM yield predictions, particularly in cases where comparison against brute-force simulation is intractable, we perform a bootstrapping procedure. Bootstrapping is a resampling technique in which multiple random samples, known as bootstrap samples, are drawn with replacement from an observed dataset. This method allows for the estimation of the sampling distribution of a statistic, to estimate uncertainties when the underlying distribution is unknown or performing error propagation is not straight-forward \cite{Efron79, Efron81}.

The typical bootstrapping procedure begins with a dataset with $N$ measurements. From this set, we construct a resampling that consists of $N$ samples from the original dataset, drawn uniformly at random with replacement. The quantity of interest is evaluated for the resampled dataset and becomes a bootstrap sample. After collecting $M$ bootstrap samples, a histogram can be constructed to show the full distribution, and the bootstrap sample mean and standard deviation can be computed. 
We perform this procedure for the MultiMSM yields, using the training trajectories as our dataset, with a small modification; we keep the number of each type of trajectory fixed during the resampling, instead of just the total number of trajectories. This is particularly important for the examples with slow monomer depletion, as a resampling that does not include enough trajectories in the lower monomer fraction regimes will give nonsensical results. 
We generate $M=1000$ bootstrap samples for each yield measurement for which we wish to estimate errors, which are reported in the main text. Fig.~\ref{SMfig::bootstrapping} shows an example histogram of bootstrapping samples for the $T=3$ capsid yield at the final simulation time, as well as the sample mean and standard deviation as a function of number of bootstrap samples. We can see the histogram is approximately normally distributed, and that the sampling estimates do not vary by much over the last $200$ samples, indicating we have performed enough bootstraps.

\subsection{Error-Based Refinement}
\label{SMsec::error_refinement}

\begin{figure}[ht!] 
\centering
    \includegraphics[width=0.75\textwidth]{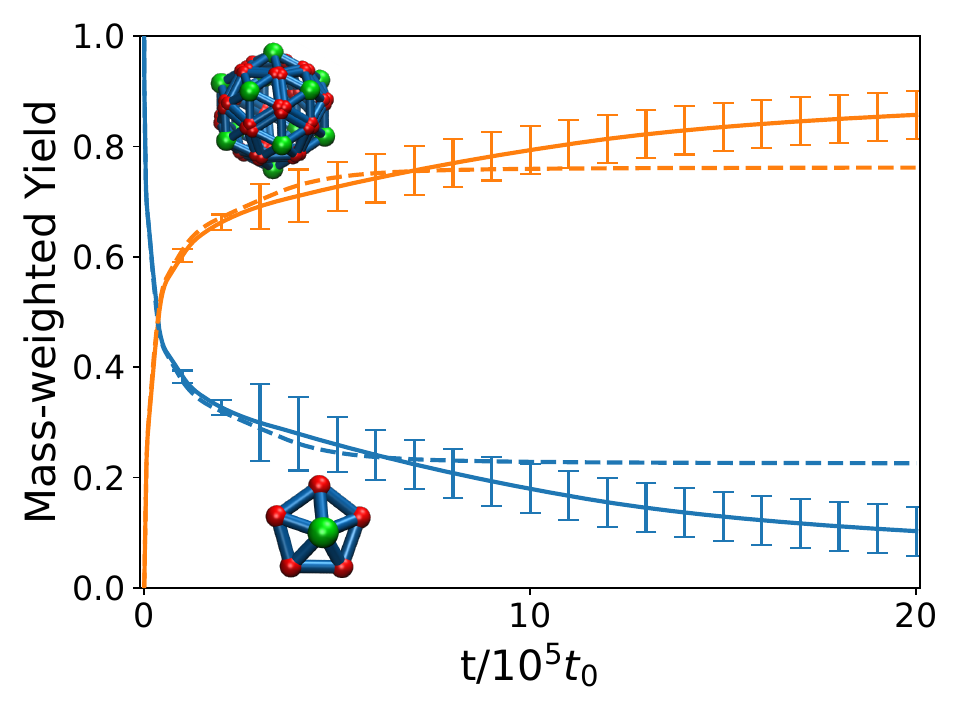}
    \caption{Example error refinement using bootstrapping error estimates. The dashed lines show the MultiMSM yield predictions for monomers (blue) and dodecahedral capsids (orange) using the converged model from the main text for $\Ebind=5.5$. Solid lines and error bars are generated using $1000$ bootstrap samples with a poor choice of monomer fraction discretization, $D_{\text{test}} = [0, 0.1, 0.35, 0.45, 0.55, 0.71, 0.81, 1.0]$. Refinement of the discretization then gives the converged MultiMSM that produced the dashed line results.
    }
    \label{SMfig::error_refinement}
\end{figure}

Bootstrapping can be performed on the MultiMSM at various time points during the assembly to estimate the model error as a function of time. These errors can then be used as a guide for refining the MultiMSM model; by identifying which monomer fraction bin the MultiMSM is using at the time the error becomes large, this gives information about where the monomer fraction discretization should be refined or more sampling should be performed. 

For example, the solid lines and error bars in Fig.~\ref{SMfig::error_refinement} were computed by bootstrapping for the dodecahedron example with binding energy $\Ebind=5.5$, but with a poor choice for the monomer fraction discretization, $D_{\text{test}} = [0, 0.1, 0.35, 0.45, 0.55, 0.71, 0.81, 1.0]$. This model differs from the converged model from the main text (dashed lines), $D_{\text{main}} = [0, 0.15, 0.25, 0.35, 0.45, 0.55, 0.71, 0.81, 1.0]$ only in the first two monomer fraction bins. 
We see a large spike in the error when the monomer fraction goes below $0.35$ for $D_{\text{test}}$, indicating an issue with the model in this bin, despite the model means being quite close at this time, which is all the optimization based refinement in Section \ref{SMsec::discretization_opt} addresses. Over longer timescales, the test model performs poorly; overestimating capsid formation and monomer depletion as well as the timescale to reach equilibrium. 

Since we have a converged and validated model with $D_{\text{main}}$, we can directly assess why the test model fails. We have noted in the main text that the critical aggregation concentration (CAC) for this example corresponds to a monomer fraction of approximately $0.25$. Slightly above this value, the system assembles slowly, while below it the system does not assemble at all. 
Both of these possibilities are captured in the discretization window $[0.1, 0.35]$ of the test model. This means that transition data above the CAC is being used to approximate the dynamics below the CAC, resulting in poor model performance. The resampling process for the bootstrap will construct models that are biased toward either side of the CAC, resulting in estimates with a large spread of possible values and therefore large error estimates.  
By breaking up this large window with the intermediate monomer fraction of $0.25$, we get the model reported in the main text, which presents far smaller error estimates and matches well with sample averages from brute-force dynamics simulations. 

While this example may seem a bit contrived, this error analysis is precisely how we arrived at our final monomer discretization for this model. Before identifying the CAC or verifying the MultiMSM predictions with simulations, the test model above was our best working model for this example, obtained partly through trial-and-error and partly by running the optimization in Section \ref{SMsec::discretization_opt} using trajectories up until a final time of $2.5 \times 10^5 \tzero$. Since there were no estimates of the late time yields at that point, we could not evaluate this test model without additional error analysis. By performing this error analysis, we were able to identify a place to refine the discretization and construct a model that accurately predicts the equilibrium yields without knowing them a priori. 
While we have not done this, an interesting possibility is to add these bootstrap error bars as a metric to the refinement objective function in Section \ref{SMsec::discretization_opt}, which would allow for model refinement in the absence of sample trajectories, the main weakness of the refinement approach we used there.

\section{MultiMSM Application Details}

The MultiMSM approach has a number of useful applications, some of which we have outlined in the main text. Here we discuss technical details of the implementations of these applications. 

\subsection{Clusterization Procedure}
\label{SMsec::clusterization}

As constructed, the MultiMSM maintains the mass-weighted normalization of cluster yields. While this is required to accurately predict bulk assembly statistics, it complicates performing calculations related to individual trajectory properties. Transition path theory calculations, such as computing the committor or the most likely assembly pathway, focus on the dynamics of individual assemblages, rather than the bulk statistics of the ensemble of clusters. 
However, the transition matrix is constructed such that if an $n$-mer loses $k < n/2$ subunits, we log a transition to the small $k$-mer in addition to the $(n-k)$-mer. This is particularly important when $k=1$, as every intermediate state would have a chance to transition to a monomer. For transition path theory calculations, we must focus on the $(n-1)$-mer instead, the larger of the split pair. 

To convert the statistics of the transition matrix calculation to the form required for calculating committor probabilities, we perform a `clusterization' procedure to the transition matrix before solving the transition path theory equations. For every state with cluster size $n>3$, we set the probability of transitioning to a monomer to zero, and instead add that probability to the most likely state with cluster size $n-1$. 
This `clusterized' transition matrix takes the place of $P_{ij}$ in Equations \eqref{eq::fcommittor}, \eqref{eq::bcommittor}, and \eqref{eq::flux}, but the standard transition matrix is still used to solve the forward Kolmogorov equation as per Eq.~\eqref{eq::FKE_multi}. 
We further generalized this procedure to remove the probability of transitioning to any state smaller than $n/2$, but the results are nearly identical to solely removing the monomer probability, likely because large cluster splittings are rare and insignificant to the dynamics. The default procedure only removes the monomer probability, but we have kept the generalized procedure as an optional flag in our clusterization method, in case it becomes necessary for other assembly systems.

\subsection{Trajectory Generation}
\label{SMsec::traj_generation}

\begin{figure}[ht!] 
\centering
    \includegraphics[width=0.99\textwidth]{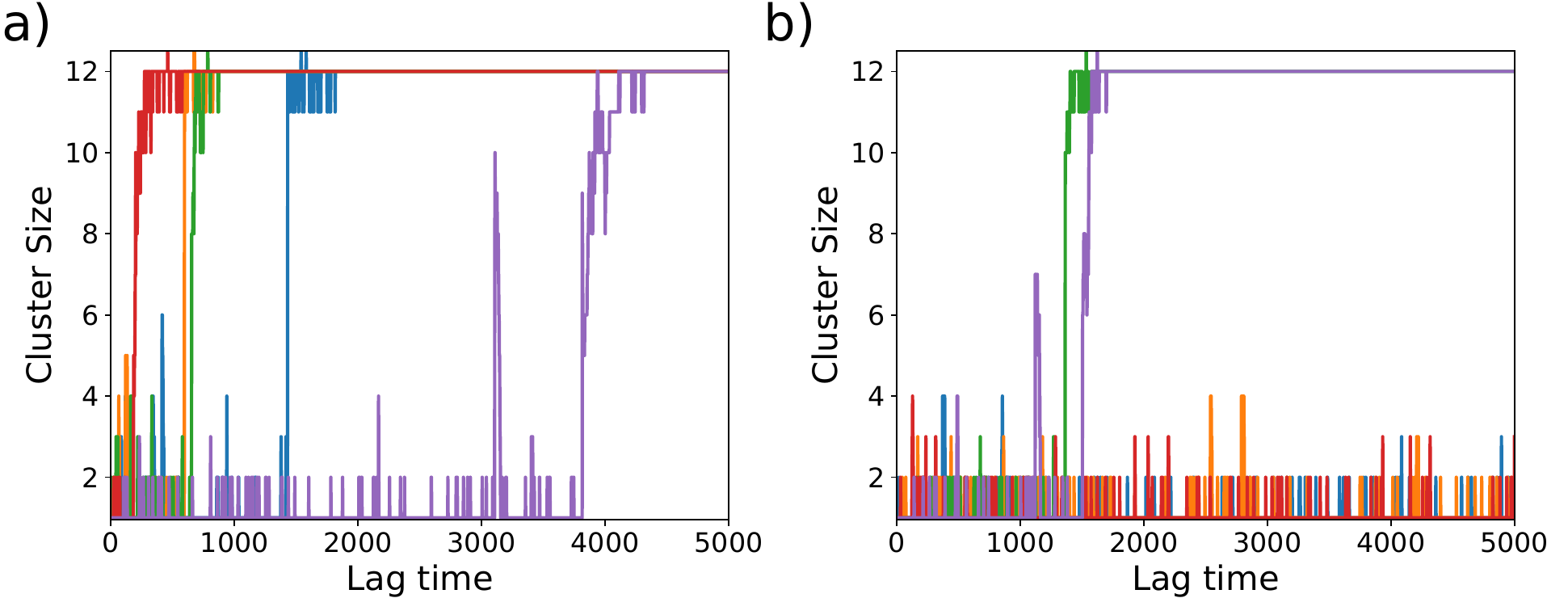}
    \caption{Examples of individual cluster assembly trajectories generated from the MultiMSM  for dodecahedral capsid assembly with $\Ebind=5.5$. Five independent trajectories were generated starting from a monomer state with initial monomer fractions 
    \textbf{(a)} $1.0$ and 
    \textbf{(b)} $0.5$. 
    The discrete state is updated every lag time, which corresponds to a simulation time of $10\tzero$. 
    }
    \label{SMfig::traj_generation}
\end{figure}

Another application of an MSM is to efficiently generate realizations of the stochastic process that it has been constructed from. For example, once an MSM has been constructed from an atomistic simulation of protein folding, it can be used to sample folding pathways with orders of magnitude less simulation time \cite{Chodera2014, Wu2018}. Such sampling can be useful to explore assembly pathways, particularly for rare events which often require large computational effort, and can be used to accelerate transition path sampling algorithms \cite{Swenson2019, Bolhuis2021}. 

In contrast to the protein folding example which focuses on a single protein, the MultiMSM models an ensemble of assembling clusters. However, as shown in the main text, it can still be used to generate assembly trajectories for individual clusters, starting from any discretized state in the model (i.e., from specific oligomers rather than unassembled monomers).

 The user must provide three properties of the trajectory; the discretized starting state, $S_0$, the starting monomer fraction, $m_0$, and the length of the trajectory, $L$, in lag times. Note that trajectory generation is inherently a single assemblage calculation, so the clusterization procedure described above must be performed here as well. 

The first step in the process is to solve the forward Kolmogorov equation for the MultiMSM, starting with all monomers, over a timescale that supports $L$ lag times after reaching monomer fraction $m_0$. Doing so will save a time-series of the monomer fraction as a function of time, as well as a time-series of references to the transition matrix used to propagate to the next time step. The next step is to choose the time step with the monomer fraction closest to $m_0$ to initialize the generation. Starting at this point, we then construct a time series of transition matrices used to update the system at every future time step. We iterate over each of these transition matrices, extract the row corresponding to the current state of the system, say $S_i$, and then sample this probability distribution to determine the next state, $S_{i+1}$. 
Note that once the forward equation has been solved once, which is the bottleneck for the generation process, the solution can be used to generate arbitrarily many trajectories with potentially different properties without being solved again, making generation extremely efficient.

Fig.~\ref{SMfig::traj_generation} shows example trajectories generated using the above procedure for dodecahedral capsid assembly with $\Ebind = 5.5$. Fig.~\ref{SMfig::traj_generation}a shows five trajectories initialized in the monomer state with a starting monomer fraction of $1.0$, i.e. the standard assembly initial condition. We see that each example assembles within $5000$ lag times ($5\times 10^4 \tzero$ simulation time) with a large spread of nucleation times.  Fig.~\ref{SMfig::traj_generation}b shows five more trajectories, again initialized in the monomer state, but with a starting monomer fraction of $0.5$. Note that the x-axis in this case represents the number of lag times since the monomer fraction first reaches $0.5$. In this case, only two of the five trajectories result in dodecahedra within $5000$ lag times, with a longer average nucleation time as one would expect at lower monomer concentrations. 

There are a few things to note about this process. First, the solution to the forward equation gives the average monomer fraction as a function of time, and we generate the next states with respect to this. This is a mean field approximation --- the protocol only explicitly simulates the cluster being sampled, so the statistics of the background monomers and intermediates are those from the MultiMSM. Thus, this protocol would not capture feedback between cluster assembly and rare, large fluctuations in the background.  Along the same lines, generating multiple trajectories must be thought of as independent realizations of the assembly from a fixed start point; these clusters will not interact with each other. However, these approximations are likely to be well justified for typical dilute solution assembly reactions.

\subsection{Free Energy Computation}
\label{SMsec::free_energy}

\subsubsection{Removal of Absorbing States}
\label{SMsec::absorbing_states}

The free energy computations detailed in the main text all involve computing the equilibrium distribution for each component transition matrix of the MultiMSM. The equilibrium distribution is given by the left eigenvector of the transition matrix corresponding to an eigenvalue of $1$. If the transition matrix is ergodic, this distribution is guaranteed to be unique and can be computed using standard numerical linear algebra techniques. This is often not the case in practice; target structures are typically designed to be stable which means they may not be sampled reversibly, particularly in cases with strong interactions between subunits. States (or sets of states) that can be entered but cannot be left are reffered to as \emph{absorbing} states, which must be identified and removed from the transition matrix before computing the equilibrium distribution. 

We use a depth-first search algorithm to determine the strongly connected component of the transition matrix; the maximal sub-graph for which every state has a non-zero probability of reaching any other state in a finite number of steps. Each state that is not a member of the strongly connected component is classified as an absorbing state. For each absorbing state, we remove its corresponding row and column in the transition matrix to form a reduced transition matrix. The rows of this reduced transition matrix need to be renormalized to sum to one, and the resulting ergodic matrix is used to compute the equilibrium distribution over the remaining states. 

\subsubsection{Equilibrium Constants}
\label{SMsec::equilibrium_constants}

Another subtlety arises when computing the equilibrium constants as a function of intermediate size. The equilibrium constants are a function of the equilibrium \emph{concentrations} of each species, not the equilibrium probability as we have computed thus far. For an $n$-mer, we can write the equilibrium concentration (in simulation units) as $\left[ n \right] = \gamma\ctot\pi_n$, where $\ctot$ is the total subunit concentration, $\pi_n$ is the total equilibrium probability for observing an $n$-mer, and $\gamma$ is a scaling factor to account for the lost concentration due to removal of absorbing states. This scaling factor is necessary because removing the absorbing states from the equilibrium calculation effectively reduces the total subunit concentration, while the probability distribution remains normalized to sum to one. Its value depends on the fraction of absorbing states, which generally increases in time, and thus depends on the monomer fraction discretization bin. For a discretization bin between monomer fractions $d_i$ and $d_{i+1}$, and corresponding times $t_i$ and $t_{i+1}$ for which $\mfrac = d_i$ and $\mfrac=d_{i+1}$, respectively, we compute $\gamma_i$ as the average yield of non-absorbing states over this time interval,
\begin{equation}
    \gamma_i = 1-\frac{1}{t_i-t_{i+1}} \sum_{j=t_{i+1}}^{t_i} p_A^j,
\end{equation}
where $p_A^j$ is the total yield of absorbing states at time $j$, computed from the original transition matrix. 

\subsubsection{Testing MultiMSM Free Energy Estimates for Dodecahedron System Subunit Dimerization}
\label{SMsec::dimerization_fe}
\begin{figure}[ht!] 
\centering
    \includegraphics[width=0.99\textwidth]{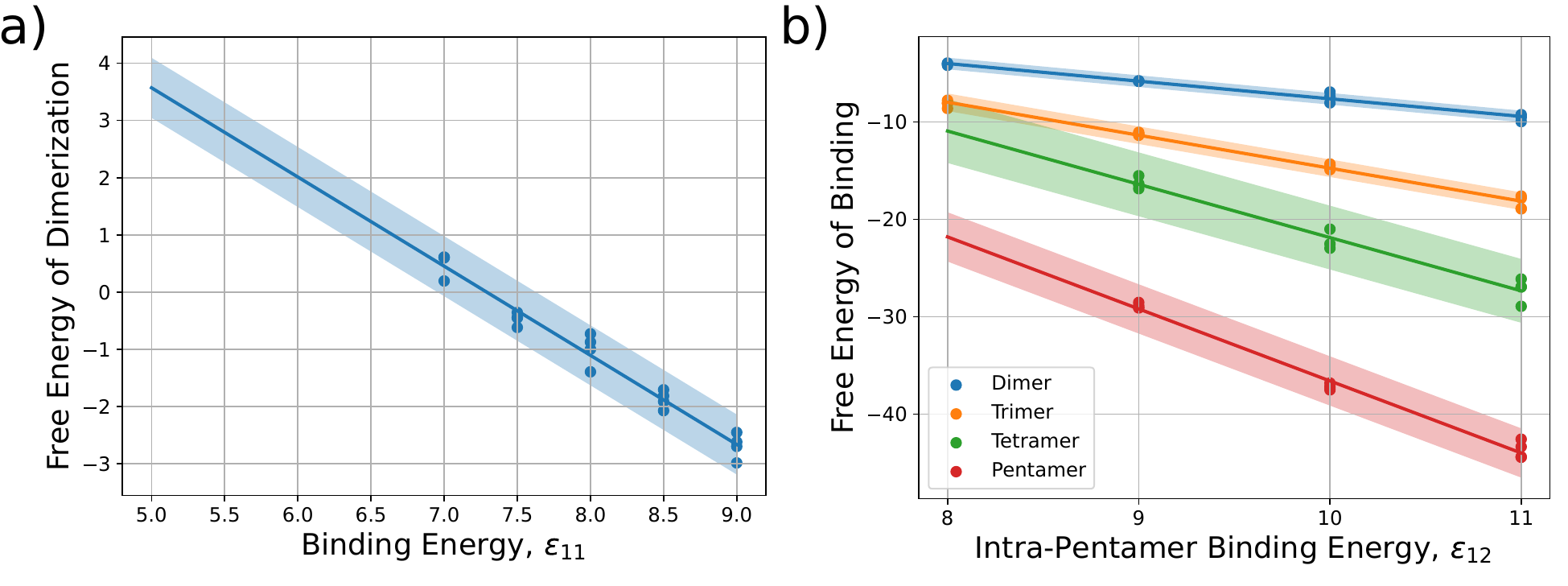}
    \caption{The binding affinity, computed directly from equilibrated simulation data, is shown as a function of the subunit-subunit binding strength for 
    \textbf{(a)} dimerization of the dodecahedron subunits and 
    \textbf{(b)} intermediate sizes up to a pentamer for the $T=3$ system. 
   Binding affinities are computed for $4$(a) and $3$(b) random seeds (points) and the line of best fit is plotted against the data. The shaded region corresponds to the standard error of the intercept for the linear regression. 
    }
    \label{SMfig::binding_free_energy_simulation}
\end{figure}

In this and the next subsection we test the free energy profile computed from the MultiMSM algorithm against independent results computed directly from Brownian dynamics simulations, respectively for the dodecahedron and $T=3$ systems. Due to the large computational cost to compute the full free energy profile as a function of intermediate size, we performed a comparison for dimerization at several binding energy values for the dodecahedron system, and for intermediates through pentamers for the $T=3$ system. 

We compute the dimerization free energy from Brownian dynamics simulations as in Refs.  \cite{Elrad2010, Hagan2011, Perlmutter2013}. We modify the subunits such that only one edge forms attractive bonds, limiting the system to only monomers and dimers. For each value of $\Ebind$, we simulate until monomer-dimer equilibrium is reached, and then average the monomer and dimer concentrations over the last $5\%$ of the equilibrated trajectory. We then compute the equilibrium constant and dimerization affinity  as $g_2 = -\kt \ln{(c_{ss}K_2)}$, where $\css$ is the same standard state reference concentration used in the MultiMSM calculation (see main text). 
We find that the free energy is approximately linear in $\Ebind$, and a least squares fit gives the expression
\begin{equation}
    g_2 = -1.56\epsilon - T s_\text{b}
\end{equation}
where $s_\text{b}/k_\text{B}=11.3$ is the binding entropy penalty. 
This linear expression can then be used to extrapolate the free energy of dimerization to the values of the binding energy used for the examples in the main text, shown in Fig.~\ref{SMfig::binding_free_energy_simulation}a. We perform this calculation $4$ times to get error bars on the linear fit and the dimerization affinities. 
Note that to apply these results to the full system (without inactivated edges) we must add the conformational entropy term $-T \Delta s_\text{c} = - \ln(5^2/2) \kt \approx -2.53 \kt$ where the numerator accounts for the $5$ distinguishable binding sites on each subunit and the denominator accounts for rotational symmetry of the dimer \cite{Hagan2011}. 

\subsubsection{Testing MultiMSM Free Energy Estimates for $T=3$ System up to Pentamers}
\label{SMsec::pentamerization_fe}

We a similar approach as in section~\ref{SMsec::dimerization_fe} to test the binding free energy for intermediates up to a pentamer for the $T=3$ subunits. In this case we perform simulations in which subunits can assemble only up to a pentamer by setting $\Ethree = 0$. We simulate this system for several values of the pentamer-bias binding energy, $\Eone$, and compute the average equilibrium concentrations for each intermediate size up to $5$ over the last $10\%$ of the trajectory. We then compute equilibrium constants and binding affinities as $g_n = -\kt \ln{(\css^{n-1}K_n)}$,  where $\css$ is the same standard state reference concentration used in the MultiMSM calculation (see main text). 
Again we find that the affinities are approximately linear with $n$, and perform a least squares fit to the data (Fig.~\ref{SMfig::binding_free_energy_simulation}b). We perform this calculation $3$ times to estimate error bars on the binding free energies at $\Eone = 11 \kt$, the value used for the example in the main text.

\subsection{Committor Calculation Verification}
\label{SMsec::committor_verification}

\begin{figure}[ht!] 
\centering
    \includegraphics[width=0.5\textwidth]{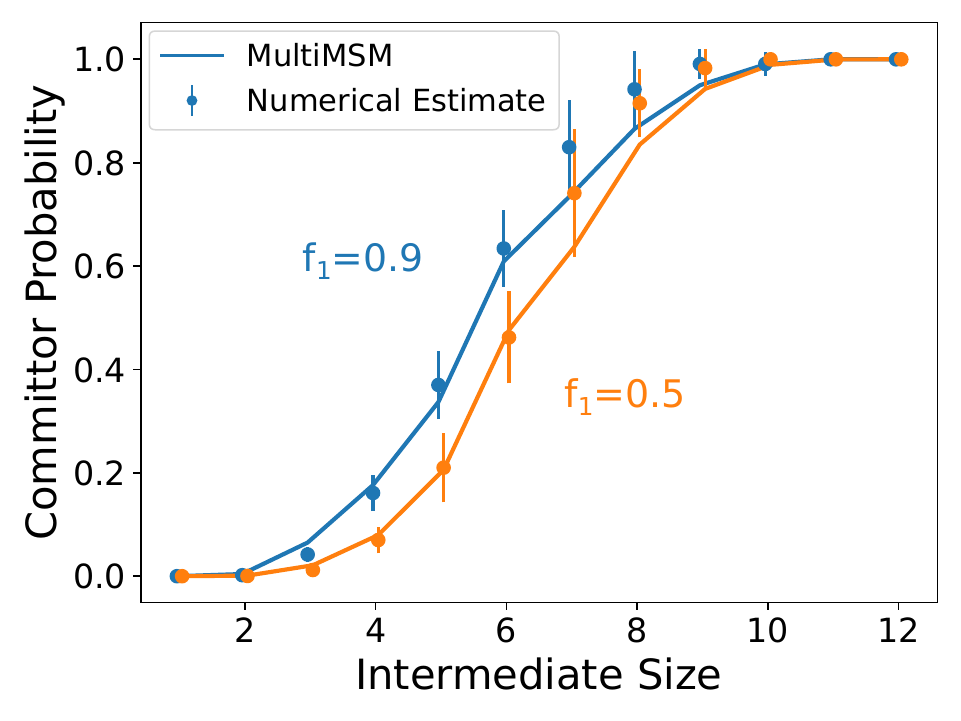}
    \caption{Comparison of the MultiMSM committor prediction with direct numerical estimates for dodecahedral capsid assembly with $\Ebind = 5.5$ at two values of the monomer fraction. The initial state is a monomer and the target state is the dodecahedral capsid. Solid lines represent the MultiMSM computed committor values at the first lag time for which the monomer fraction reaches the labeled value. Points and error bars are computed directly from the trajectories used to construct the MultiMSM. 
    }
    \label{SMfig::committor_verification}
\end{figure}

While the MultiMSM committor predictions qualitatively match our expectations, we perform a quantitative verification against direct numerical estimates to validate our approach and clusterization procedure. We use the MultiMSM training data as our source for transitions. 
To estimate the committors numerically, we identify every instance of a cluster trajectory reaching a given intermediate size within a monomer fraction window of $\pm 0.05$ of a desired monomer fraction value. If the trajectory reaches the target before returning to the monomer state, we increment our estimator and number of samples by $1$. If the trajectory goes back to the monomer state first,, we increment our number of samples by $1$ without incrementing the estimator. If the trajectory does not reach either state, we exclude it from the calculation. Dividing the estimator by the number of samples gives an estimate of the committor probability for each intermediate size. 
We estimate error bars by splitting our trajectories into $6$ batches of roughly equal size and taking the standard deviation over each batches' sample mean. 

Fig.~\ref{SMfig::committor_verification} shows a comparison of two committor results in the main text for dodecahedral capsid assembly with $\Ebind=5.5$ with our direct numerical estimates, for two values of the monomer fraction. 
There is good agreement between the MultiMSM prediction and numerical estimates in both cases. The MultiMSM seems to slightly overestimate the committor for smaller intermediate sizes and underestimate it for larger intermediate sizes. 
The agreement is worst for intermediate sizes between $7{-}9$ for both examples. This is likely due to limited sampling of intermediates of this size; smaller intermediates are more likely to be sampled several times per trajectory since they more readily dissociate into monomers whereas intermediates larger than the critical nucleus are often sampled only once per trajectory. This affects both the MultiMSM estimate as well as the numerical estimate, which has the largest error bars at these values.

\subsection{Additional MultiMSM Verification}
\label{SMsec::additional_verification}
\begin{figure}[ht!] 
\centering
    \includegraphics[width=0.5\textwidth]{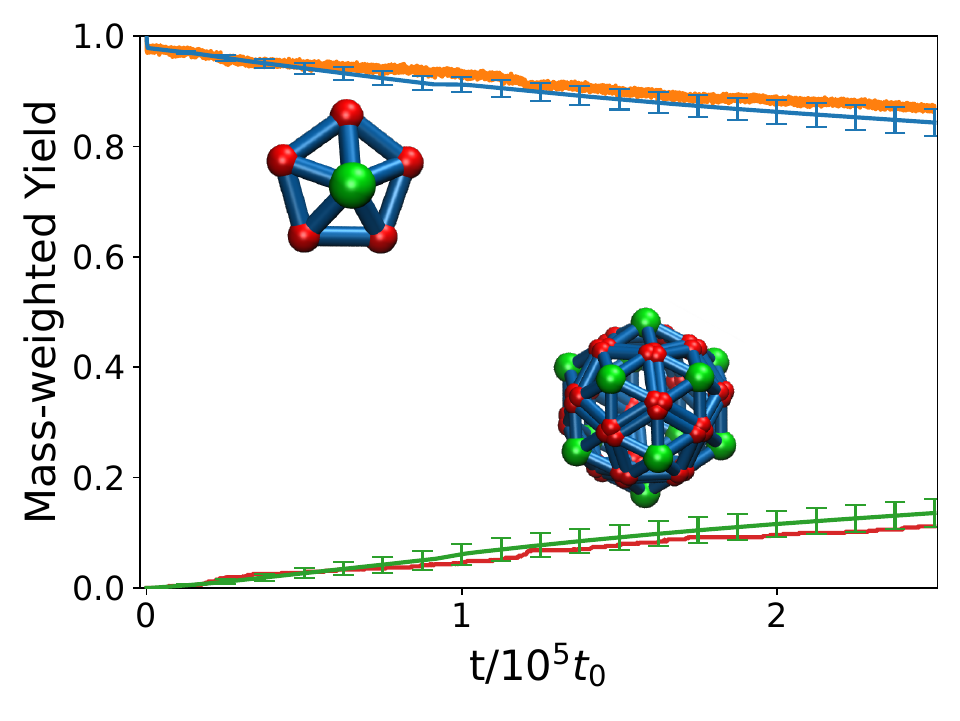}
    \caption{MultiMSM predictions (smooth curves with error bars) for dodecahedron capsid assembly dynamics for $\Ebind=5$ and $\ctot = 0.0156$ (same parameters as Fig.~\ref{fig::dodec_verify}c in the main text) compared to results from brute-force dynamics simulations (noisy curves) over the accessible simulation timescale, $\TF = 2.5 \times 10^5 \tzero$. Error bars are estimated for the MultiMSM by bootstrapping with $1000$ resamplings (see Section \ref{SMsec::bootstrapping}).
    }
    \label{SMfig::dodec_E5_verify}
\end{figure}

The verification of the MultiMSM prediction of dodecahedral capsid assembly for $\Ebind=5$ over the computationally accessible timescale was omitted from Fig.~\ref{fig::dodec_verify}c due to the large difference in this timescale compared to reaching steady state. The result is instead shown here in Fig.~\ref{SMfig::dodec_E5_verify}, comparing the monomer and capsid yields predicted by the MultiMSM with sample-averaged estimates from $50$ brute-force dynamics simulations. We again see good agreement over this timescale, with our brute force estimates generally within the error bars of the MultiMSM prediction.

\end{document}